\documentclass[12pt]{article}
\usepackage[utf8]{inputenc}
\usepackage{amsmath}
\usepackage{amssymb}
\usepackage{bbold}
\usepackage[title]{appendix}
\usepackage{comment}
\usepackage{float}
\usepackage{array}
\usepackage{hyperref}
\usepackage{natbib}
\usepackage{graphicx}
\usepackage{xcolor}
\usepackage{appendix}

\newcommand{\newc}{\newcommand}
\newc{\bc}{\begin{center}}
\newc{\ec}{\end{center}}
\newc{\be}{\begin{enumerate}}
\newc{\ee}{\end{enumerate}}
\newc{\bi}{\begin{itemize}}
\newc{\ei}{\end{itemize}}
\newc{\bd}{\begin{description}}
\newc{\ed}{\end{description}}
\newc{\und}[1]{\underline{#1}}
\newc{\E}{\mbox{E}}
\newc{\V}{\mbox{V}}
\newc{\N}{\mbox{N}}
\newc{\B}{\mbox{Bin}}
\newc{\Bern}{\mbox{Bern}}
\newc{\Po}{\mbox{Po}}
\newc{\IG}{\mbox{IG}}
\newc{\Gam}{\mbox{Gam}}
\newc{\bdP}{\mbox{P}}
\newc{\bdp}{\mathsf{p}}
\newc{\bdphat}{\hat{\mathsf{p}}}
\newc{\odds}{\mbox{odds}}
\newc{\OR}{\mbox{OR}}
\newc{\stderr}{\mbox{s.e.}}
\newc{\logit}{\mbox{logit}}
\newc{\sign}{\mbox{sign}}
\newc{\SD}{\mbox{SD}}
\newc{\bdmu}{\mbox{\boldmath $\mu$}}
\newc{\bdSigma}{\mbox{\boldmath $\Sigma$}}
\newc{\bdLambda}{\mbox{\boldmath $\Lambda$}}
\newc{\bdmuhat}{\mbox{\boldmath $\hat{\mu}$}}
\newc{\bdeta}{\mbox{\boldmath $\eta$}}
\newc{\bdtheta}{\mbox{\boldmath $\theta$}}
\newc{\bdbeta}{\mbox{\boldmath $\beta$}}
\newc{\bdgamma}{\mbox{\boldmath $\gamma$}}
\newc{\bdetahat}{\mbox{\boldmath $\hat{\eta}$}}
\newc{\bdbetahat}{\mbox{\boldmath $\hat{\beta}$}}
\newc{\bdgammahat}{\mbox{\boldmath $\hat{\gamma}$}}
\newc{\bdthetahat}{\mbox{\boldmath $\hat{\theta}$}}
\newc{\bdvareps}{\mbox{\boldmath $\varepsilon$}}
\newc{\bdzero}{\mbox{\boldmath $0$}}
\newc{\bdone}{\mbox{\boldmath $1$}}
\newc{\bdnu}{\mbox{\boldmath $\nu$}}
\newc{\bdell}{\mbox{\boldmath $\ell$}}
\newc{\bdxi}{\mbox{\boldmath $\xi$}}
\newc{\bdomega}{\mbox{\boldmath $\omega$}}
\newc{\bdepsilon}{\mbox{\boldmath $\varepsilon$}}
\newc{\bdI}{\mathbf{I}}
\newc{\bdX}{\mbox{\boldmath $X$}}
\newc{\bdA}{\mbox{\boldmath $A$}}
\newc{\bdB}{\mbox{\boldmath $B$}}
\newc{\bdC}{\mbox{\boldmath $C$}}
\newc{\bdD}{\mbox{\boldmath $D$}}
\newc{\bdG}{\mbox{\boldmath $G$}}
\newc{\bdJ}{\mbox{\boldmath $J$}}
\newc{\Ktil}{\tilde{K}}
\newc{\Khat}{\hat{K}}
\newc{\bda}{\mbox{\boldmath $a$}}
\newc{\bdb}{\mbox{\boldmath $b$}}
\newc{\bdc}{\mbox{\boldmath $c$}}
\newc{\bde}{\mbox{\boldmath $e$}}
\newc{\bdu}{\mbox{\boldmath $u$}}
\newc{\bdv}{\mbox{\boldmath $v$}}
\newc{\bdx}{\mbox{\boldmath $x$}}
\newc{\bdy}{\mbox{\boldmath $y$}}
\newc{\bdz}{\mbox{\boldmath $z$}}
\newc{\bdr}{\mbox{\boldmath $r$}}
\newc{\bdQ}{\mbox{\boldmath $Q$}}
\newc{\bdR}{\mbox{\boldmath $R$}}
\newc{\bdY}{\mbox{\boldmath $Y$}}
\newc{\bdT}{\mbox{\boldmath $T$}}
\newc{\bdW}{\mbox{\boldmath $W$}}
\newc{\bdWtil}{\tilde{\mbox{\boldmath $W$}}}
\newc{\bdH}{\mbox{\boldmath $H$}}
\newc{\bdL}{\mbox{\boldmath $L$}}
\newc{\bdU}{\mbox{\boldmath $U$}}
\newc{\bdV}{\mbox{\boldmath $V$}}
\newc{\Multinom}{\mbox{Multinom}}
\newc{\Var}{\mbox{Var}}
\newc{\var}{\mbox{var}}
\newc{\diag}{\mbox{diag}}
\newc{\thetahat}{\hat{\theta}}
\newc{\tr}{\mbox{tr}}
\newc{\phat}{\hat{p}}
\newc{\Xbar}{\bar{X}}
\newc{\xbar}{\bar{x}}
\newc{\Ybar}{\bar{Y}}
\newc{\ybar}{\bar{y}}
\newc{\dbar}{\bar{d}}
\newc{\yhat}{\hat{y}}
\newc{\bdyhat}{\mbox{\boldmath $\hat{y}$}}
\newc{\ytil}{\tilde{y}}
\newc{\ftil}{\tilde{f}}
\newc{\Ho}{\mbox{\bf H}_o}
\newc{\Ha}{\mbox{\bf H}_a}
\newc{\phatYX}{\phat_Y - \phat_X}
\newc{\SSG}{\mbox{SSG}}
\newc{\SSB}{\mbox{SSB}}
\newc{\SSE}{\mbox{SSE}}
\newc{\SST}{\mbox{SST}}
\newc{\SSR}{\mbox{SSR}}
\newc{\SSAB}{\mbox{SSAB}}
\newc{\MSG}{\mbox{MSG}}
\newc{\MSB}{\mbox{MSB}}
\newc{\MSE}{\mbox{MSE}}
\newc{\MST}{\mbox{MST}}
\newc{\MSAB}{\mbox{MSAB}}
\newc{\dfE}{\mbox{dfE}}
\newc{\dfG}{\mbox{dfG}}
\newc{\dfB}{\mbox{dfB}}
\newc{\dfT}{\mbox{dfT}}
\newc{\dfAB}{\mbox{dfAB}}
\newc{\muhat}{\hat{\mu}}
\newc{\betahat}{\hat{\beta}}
\newc{\alphahat}{\hat{\alpha}}
\newc{\etahat}{\hat{\eta}}
\newc{\phihat}{\hat{\phi}}
\newc{\sigmahat}{\hat{\sigma}}
\newc{\cl}{\centerline}
\newc{\R}{\mathbb{R}}
\newc{\trans}{^\mathsf{T}}
\newc{\xtx}{\bdX\trans\bdX}
\newc{\xxtxx}{\bdX(\xtx)^{-1}\bdX\trans}
\newc{\argmin}{\operatornamewithlimits{argmin}}
\newc{\argmax}{\operatornamewithlimits{argmax}}
\newc{\incg}[2]{
\includegraphics[width=#1\textwidth]{#2}
}

\title{Rating competitors in games with strength-dependent tie probabilities}
\author{Mark E.\ Glickman\thanks{Address for correspondence: 
Department of Statistics, Harvard University, 1 Oxford Street,
Cambridge, MA  02138.  Email address: {\tt glickman@fas.harvard.edu}} \\ 
Department of Statistics\\
Harvard University
}
\date{}

\begin{document}

\maketitle

\addtolength{\baselineskip}{12pt}

\begin{abstract}
Competitor rating systems for head-to-head games are typically used to 
measure playing strength from game outcomes.  
Ratings computed from these systems are often used to select top 
competitors for elite events, for pairing players of similar strength 
in online gaming, and for players to track their own strength over time.  
Most implemented rating systems assume only win/loss outcomes, and treat 
occurrences of ties as the equivalent to half a win and half a loss.  
However, in games such as chess, the probability of a tie (draw) is 
demonstrably higher for stronger players than for weaker players, so that 
rating systems ignoring this aspect of game results may produce strength 
estimates that are unreliable.  
We develop a new rating system for head-to-head games based on a
model by \citet{glickman2025paired} that explicitly 
acknowledges that a tie may 
depend on the strengths of the competitors.  
The approach uses a Bayesian dynamic modeling framework. 
Within each time period, 
posterior updates are computed in closed form using a single Newton-Raphson 
iteration evaluated at the prior mean.
The approach is demonstrated on a large dataset of chess games played in 
International Correspondence Chess Federation tournaments. 
\end{abstract}


\section{Introduction}\label{sec:intro}

Rating systems for chess play a vital role in organizing tournaments 
and matches, tracking individual progress, and facilitating fair matchmaking. 
They provide a standardized framework to compare the strength of players based 
on game outcomes, enabling national and international
federations to assign titles, seed tournaments, 
and monitor competitive balance across player pools. 
As competitive chess continues to integrate into global and online formats, 
the need for dynamic, data-driven rating systems has never been greater.

Numerous systems have been proposed to address the need for reliable, 
scalable rating methods in chess and other competitive settings.
Over the decades, several chess rating systems have 
gained widespread adoption.
Among the most influential is the system developed by
\citet{elo1978rating}, introduced in the late 1950s and implemented
by the US Chess Federation in the early 1960s. 
It was subsequently adopted by the
World Chess Federation (FIDE) in the 1970s.
Elo's approach is loosely based on the model of
\citet{bradley_rank_1952}, a foundational probability
model for paired comparisons.
Building on this framework,
the Glicko \citep{glickman1999parameter} and
Glicko-2 \citep{glickman2001dynamic} systems were introduced in
the 1990s and early 2000s. 
These models approximate
Bayesian state-space frameworks in which 
game outcomes are governed by the Bradley-Terry model 
and player abilities evolve according to Gaussian processes.
Today, Elo-based and Glicko-style systems
are widely used by national and international organizations, 
online platforms, and competitive gaming leagues, 
serving as essential tools for assessing
player strength in large-scale environments.
A comprehensive review 
of rating systems currently in use can be
found in \citet{glickman2024models}.

The International Correspondence Chess Federation (ICCF), the principal
international body for organizing competitive correspondence chess, 
has long relied on a rating system based largely on Elo's formulation. 
However, the system has shown critical shortcomings when applied to the 
distinctive features of ICCF competition. 
Chief among these is the exceptionally high rate of drawn games among 
top-level players. 
The Elo, Glicko, and Glicko-2 systems do not explicitly model the probability 
of a draw; instead, they treat a draw as equivalent in informational value to 
the average of a win and a loss. 
While \citet{szczecinski2020understanding} introduced a 
modification of the Elo framework, building on the model 
by \citet{Davidson1970On}, 
to incorporate draws explicitly,
this extension assumes a constant draw frequency and does not allow the 
probability of a draw to vary with player strength. 
In ICCF play, approximately 95\% of games between top-rated players end in a 
draw, whereas lower-rated players draw far less frequently.
This strength-dependent draw behavior is not adequately captured by 
rating systems that lack a mechanism for modeling tie probabilities as a 
function of ability. 
As a result, these systems may produce stagnant ratings, especially for elite
players whose results are overwhelmingly drawn.
Recognizing these structural limitations, ICCF officials initiated efforts
to reevaluate their rating methodology and explore alternative systems better 
suited to the unique dynamics of correspondence chess.

This paper introduces a new rating system designed to address 
limitations of existing methods in modeling draw behavior.
The proposed system 
builds on the recently developed model by 
\citet{glickman2025paired},
which explicitly models the
probability of a draw as a function of 
player strength.
In addition, it incorporates 
a dynamic framework that 
allows player ratings to evolve over time.
Player abilities are modeled as following 
a normal random walk,
capturing natural fluctuations in strength, and 
a Bayesian filtering procedure is used to update ratings based on recent 
game outcomes. 
The system balances the need to incorporate historical performance with 
the practical goal of emphasizing current ability. 
By more accurately modeling draw tendencies and supporting
time-sensitive updates, the method provides a principled and scalable 
solution for rating players in correspondence chess and related
competitive settings.
The ICCF officially adopted and implemented this rating system in 2023.

This paper is organized as follows.
Section~\ref{sec:background} provides background on the challenges
faced by the ICCF, including an empirical demonstration of the limitations
in their implementation of the Elo rating system.
In Section~\ref{sec:model},
we introduce a fully Bayesian state-space model for game outcomes, 
building on the framework developed in \citet{glickman2025paired}.
The model
accounts for strength-dependent draw probabilities,
incorporates the possibility that stronger players may better
exploit the advantage of playing white,
and allows player abilities to evolve over time.
In Section~\ref{sec:filter}, we present an 
approximate filtering algorithm that translates a one-step
Bayesian updating process into a computationally efficient method for
real-time rating updates.
Section~\ref{sec:application} applies the proposed approach to 
ICCF game outcomes from 2016 to 2022. 
We conclude
in Section~\ref{sec:discuss} with a discussion of findings and implications.

\section{ICCF background}\label{sec:background}

The ICCF, recognized by FIDE as the official global authority for correspondence chess, 
was established in 1951. 
It oversees international correspondence competitions conducted via postal mail and, 
more recently, through an online server platform. 
The ICCF maintains a structured system of player titles and ratings, 
closely aligned with those used in over-the-board play. 
Unlike traditional chess tournaments, correspondence games unfold over extended periods, 
often allowing weeks or even months per move. 
Players are permitted to use computational resources such as databases, 
opening references, and chess engines, 
although engine use is regulated in certain events. 
The ICCF organizes both individual and team championships, 
including world title events. 
With the transition from postal to digital formats in recent decades, 
the ICCF now facilitates play primarily through its online platform, 
enabling thousands of players worldwide to engage in structured, 
asynchronous international competition.

In 2021, I was invited by the ICCF to develop a new rating system to replace 
its long-standing Elo-based framework. 
The \citet{elo1978rating} system updates a player's rating,
an estimate of their playing strength,
based on game outcomes
using relatively simple computations.
A player’s rating increases after a win or a draw against a higher-rated opponent, 
and decreases after a loss or a draw against a lower-rated opponent. 
Elo ratings are scaled so that a rating of 1500 corresponds to
average strength, ratings of 1000 or below indicate relatively weak 
players, and ratings above 2000 correspond to expert-level strength.
Ratings of 2500 and higher are typically associated with world-class players.
If $R_i$ and $R_j$ denote the Elo ratings of players $i$ and $j$, respectively,
the expected score, or ``winning expectancy'' in Elo's terminology, 
for player $i$ is given by the formula
\begin{equation}
\frac{1}{1 + 10^{-(R_i-R_j)/400}},
\label{eq:elo-we}
\end{equation}
which is a rescaled inverse-logit function of
the rating difference.

The ICCF’s initiative was motivated by well-documented concerns about the limitations 
of its existing rating system, particularly its inability to accommodate the 
distinctive features of correspondence chess. 
Foremost among these concerns was the extremely high frequency of drawn games at 
the highest levels of play. 
A primary factor contributing to this pattern is the widespread use of powerful 
chess engines by top players to support their analysis. 
Given the current capabilities of these engines, it is exceedingly difficult for even 
the strongest players to secure a meaningful advantage when both sides are making 
near-optimal moves. 
Consequently, the Elo-based system struggled to register informative rating changes 
from drawn outcomes and failed to account for how the value of a draw might vary with 
the relative strength of the opponents. 
This limitation resulted in rating stagnation and reduced differentiation among 
elite players, underscoring the need for a new system better suited to the realities 
of modern correspondence chess.

The relationship between draw frequency and player strength in ICCF competition is 
evident from an empirical analysis. 
We examined all ICCF games completed between 2016 and 2022 in which the pre-game 
rating difference between the two players was exactly 50 points. 
These pre-game ICCF ratings were used as proxies for player strength, 
acknowledging that the broader objective of this work is to develop a more 
refined rating algorithm. 
A 50-point difference corresponds to a winning expectancy of approximately 0.57
for the higher-rated player, 
reflecting matchups between players of similar ability.
This restriction yielded a dataset of 149,731 game outcomes.
To explore the relationship between draw probability and player strength, 
we fit a logistic regression model for the probability that a game resulted 
in a draw (as opposed to a decisive outcome), using generalized additive modeling 
\citep{hastie1986generalized,wood2011} implemented in the {\tt mgcv} package in R. 
The model included smoothing spline terms for both the rating difference and the 
average rating of the two players.

\begin{figure}[ht]
\centerline{
\includegraphics[width=0.9\textwidth]{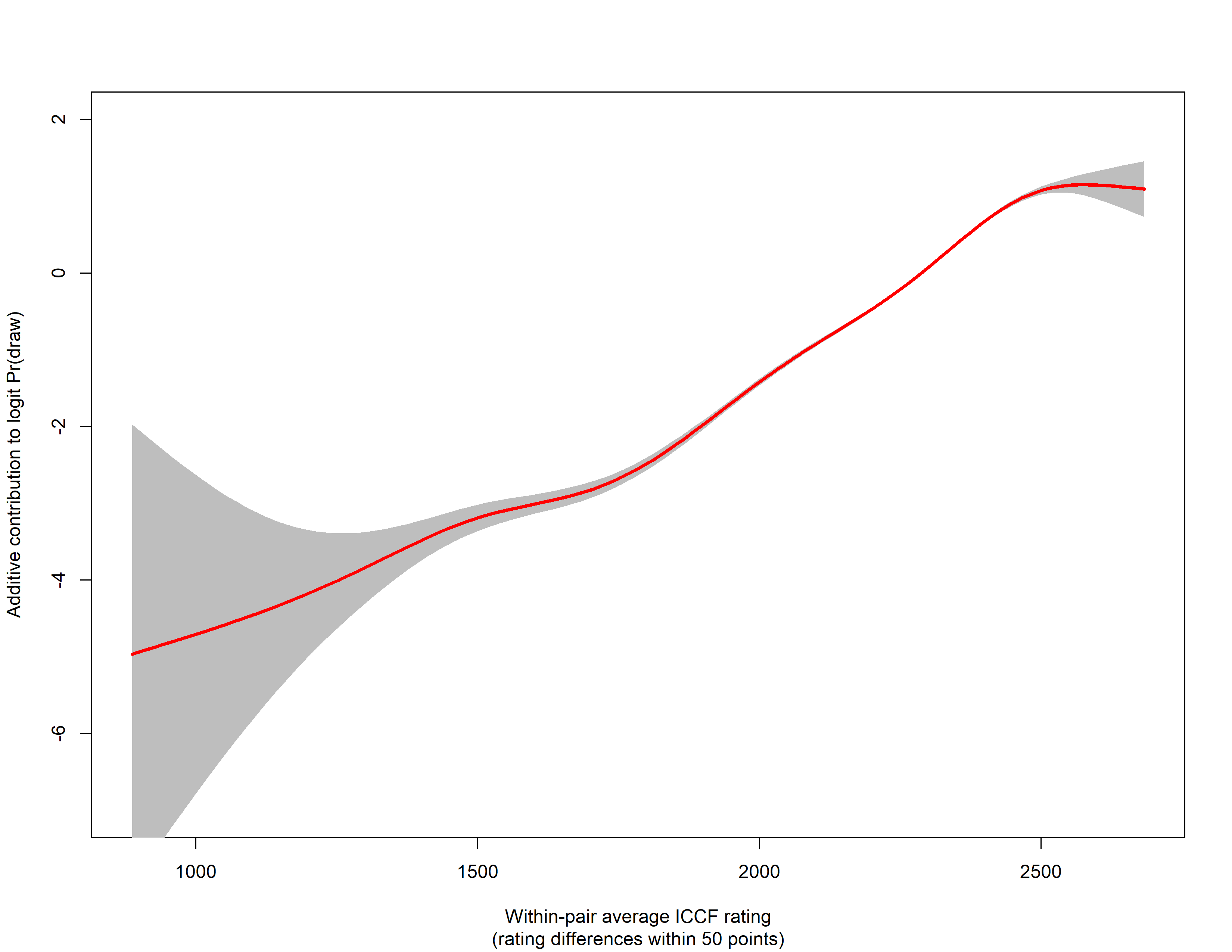}
}
\caption{
Log-odds of higher-rated player
drawing as a function of the average rating between two 
players, controlling for rating difference.
The shaded area represents 95\% pointwise confidence limits.
}
\label{fig:draw-ratavg}
\end{figure}

Figure~\ref{fig:draw-ratavg} illustrates the relationship between the 
average ICCF rating of two players 
and the additive contribution to the log-odds of a game ending in a draw. 
The red curve represents the estimated smooth effect, while the shaded 
region indicates a 95\% pointwise confidence band. 
The plot shows a strong positive linear association: 
as the average player strength increases, the log-odds of a draw also increases. 
This implies that higher-rated players, those more likely to use strong 
engine-assisted analysis, are significantly more prone to drawing games. 
The effect appears to level off slightly at the highest rating levels, 
suggesting a saturation point where draws are nearly inevitable.

\section{A Bayesian state-space model for paired comparisons}\label{sec:model}

The development of a revised rating system for the ICCF begins with the 
specification of a probability model for game outcomes. 
We adopt a Bayesian framework, consistent with the foundations of the Glicko and 
Glicko-2 rating systems. 
Our starting point is the model proposed by \citet{glickman2025paired}, 
which explicitly accounts for the probability of a draw as a function of player strength.

Consider $M$ chess players in a league, and
divide time into $T$ equally spaced periods. 
Let $\theta_{1t},\ldots, \theta_{Mt}$
denote the latent strength parameters of the players at time period $t$,
for $t=1,\ldots, T$.
For a game between players $i$ and $j$ in period $t$, define the game outcome variable as
\begin{equation}
\label{eq:y_def}
\begin{aligned}
Y_{ijt} = \left\{
  \begin{array}{rl}
  0 & \mbox{if player $i$ loses to opponent $j$ in period $t$,}\\
  1/2 & \mbox{if player $i$ ties/draws opponent $j$ in period $t$,}\\
  1 & \mbox{if player $i$ defeats opponent $j$ in period $t$.}
  \end{array} \right.
\end{aligned}
\end{equation}
Following \citet{glickman2025paired}, we model the probabilities of each
possible outcome as
\begin{equation}
\bdP(Y_{ijt} = y) =
\begin{cases}
\displaystyle \frac{\exp\left(\theta_{it} + x_{ijt}(\alpha_0 + \alpha_1 \cdot \frac{\theta_{it} + \theta_{jt}}{2})/4 \right)}{S_{ijt}} & \text{if } y = 1, \\*[12pt]
\displaystyle \frac{\exp\left(\theta_{jt} - x_{ijt}(\alpha_0 + \alpha_1 \cdot \frac{\theta_{it} + \theta_{jt}}{2})/4 \right)}{S_{ijt}} & \text{if } y = 0, \\*[12pt]
\displaystyle \frac{\exp\left(\beta_0 + (1 + \beta_1) \cdot \frac{\theta_{it} + \theta_{jt}}{2} \right)}{S_{ijt}} & \text{if } y = 1/2.
\end{cases}
\label{eq:novel}
\end{equation}
% 
% \begin{equation}
% \begin{aligned}
% \bdP(Y_{ijt}=1) =  &
% \dfrac{
% \exp\left(\theta_{it} + x_{ijt}\left( \alpha_0 + \alpha_1 \cdot \frac{\theta_{it}+\theta_{jt}}{2} \right) /4\right)
% }{S_{ijt}} \\*[12pt]
% \bdP(Y_{ijt}=0) =  &
% \dfrac{
% \exp\left(\theta_{jt} - x_{ijt}\left( \alpha_0 + \alpha_1 \cdot \frac{\theta_{it}+\theta_{jt}}{2} \right) /4\right)
% }{S_{ijt}} \\*[12pt]
% \bdP(Y_{ijt}=\frac{1}{2}) =  &
% \dfrac{
% \exp\left( \beta_0 + (1+\beta_1) \cdot \frac{\theta_{it}+\theta_{jt}}{2} \right)
% }{S_{ijt}}
% \end{aligned}
% \end{equation}
where  $x_{ij}=1$ if player $i$ has the white pieces and $-1$ if playing black,
and where
$\alpha_0$, $\alpha_1$, $\beta_0$ and $\beta_1$ are real-valued parameters 
shared across all player pairs. 
The normalizing constant $S_{ijt}$ is the sum of the numerators,
\begin{eqnarray}\label{eq:denom}
S_{ijt} &=&
   \exp\left(\theta_{it} + x_{ijt}\left( \alpha_0 + \alpha_1 \cdot \frac{\theta_{it}+\theta_{jt}}{2} \right) /4\right) +
\exp\left(\theta_{jt} - x_{ijt}\left( \alpha_0 + \alpha_1 \cdot \frac{\theta_{it}+\theta_{jt}}{2} \right) /4\right) \nonumber \\
&&+
\exp\left( \beta_0 + (1+\beta_1) \cdot \frac{\theta_{it}+\theta_{jt}}{2} \right).
\end{eqnarray}

Several observations about the model in~(\ref{eq:novel}) are worth highlighting.
When $\alpha_1=\beta_1 = 0$, the model reduces to a
non-dynamic form proposed by \citet{david_method_1988}.
In this special case, the parameter
$\alpha_0$ captures a fixed advantage for the player with the white pieces: 
a positive value indicates an advantage for white, while a negative value 
indicates an advantage for black. 
The parameter $\beta_0$ governs the overall
frequency of draws, with larger values corresponding
to higher draw probabilities.
The model in~(\ref{eq:novel}) 
extends that of \citet{david_method_1988} by allowing 
both the draw probability and the white advantage to vary as functions 
of player strength. 
Specifically, when $\alpha_1>0$, the probability that white wins increases 
with the average strength of the two players,
controlling for their strength difference.
Similarly, when $\beta_1>0$, the probability of a draw
increases with the average player strength.
Further details on this model and its implications can be found
in~\citet{glickman2025paired}.

To account for changes in player ability over time, we incorporate a dynamic 
component into the 
model via a stochastic innovation process. 
Specifically, each player's latent strength is assumed to evolve
according to  a normal random walk across discrete time periods. 
For $t=1,\ldots, T-1$,
\begin{equation}
    \begin{aligned}
        \theta_{i,t+1} \sim \N(\theta_{it}, \tau^2),
    \end{aligned}
    \label{eq:innov}
\end{equation}
where $\tau^2$ is the innovation variance 
governing the variability in strength between consecutive periods. 
This specification assumes nonstationarity in player strength, 
with variability accumulating over time.
This formulation allows player abilities to fluctuate incrementally over 
time, capturing natural changes due 
to factors such as practice, aging, or varying levels of competitive engagement. 
By embedding this stochastic structure, the model
remains responsive to recent performance while preserving reasonable estimates 
for players with sparse or infrequent activity.

This modeling strategy aligns with prior work on 
time-varying strength estimation, including methods proposed by 
\cite{fahrmeir1994dynamic}, \citet{glickman1999parameter},
and \citet{ingram2021extend},
which employ stochastic processes to capture longitudinal variation in 
latent ability. 
The normal random walk assumption introduces temporal smoothness,
helping to avoid overfitting to short-term fluctuations while still capturing 
meaningful trends in player performance. 
Although autoregressive formulations could offer more structured dynamics and 
the advantage of a stationary distribution, 
they pose challenges, particularly when players have no observed outcomes in a
given period, leading to parameter drift.
These limitations make the normal random walk a more robust and practical 
choice for modeling evolving player strength over time.

The Bayesian modeling formulation is completed by specifying
a prior distribution for the strength parameters at time $t=1$. 
For each player $i$, the initial latent strength is assumed to follow 
a normal distribution with player-specific mean $\mu_{i1}$ and
variance $\sigma_{i1}^2$, 
\begin{equation}
 \begin{aligned}
     \theta_{i1} \sim \N(\mu_{i1}, \sigma_{i1}^2),
 \end{aligned}
 \label{eq:prior}
\end{equation}
for $i=1,\ldots, M$.
This prior encodes baseline beliefs about each player's 
ability before any game outcomes are observed. 
When historical data or prior ratings are available, such as from earlier
rating systems or previous competitions, they can inform the choice of 
$\mu_{i1}$ and $\sigma_{i1}^2$, grounding the model in established assessments 
of skill.
For experienced players with extensive performance histories, smaller 
values of $\sigma_{i1}^2$ reflect greater confidence in their prior ratings.
In contrast, for new players with no prior games recorded, the prior 
parameters may be set using default values based on demographic 
characteristics or assumed to reflect the broader population distribution 
of playing strengths.

The combination of the three model components in~(\ref{eq:novel}), 
(\ref{eq:innov}) and (\ref{eq:prior})
defines a dynamic generalized linear model
\citep{west1985dynamic},
a specific instance of a broader class of state-space models \citep{durbin2012time}.
The conditional posterior distribution of the strength parameters, 
given game outcomes
$\bdy$ over all $T$ time periods and the non-strength parameters, 
is given by
\begin{eqnarray}
\lefteqn{
p(\bdtheta_1,\ldots,\bdtheta_T \mid \alpha_0,\alpha_1,\beta_0,\beta_1,\tau^2,\bdy)
}\nonumber \\
&\propto&
p(\bdtheta_1)
\prod_{t=1}^T p(\bdy_t \mid \bdtheta_t,\alpha_0,\alpha_1,\beta_0,\beta_1)
\prod_{t=2}^T p(\bdtheta_t \mid \bdtheta_{t-1},\tau^2),
\label{eq:posterior}
\end{eqnarray}
where $\bdy_t$ denotes the vector of game results during period $t$,
and $\bdtheta_t = (\theta_{1t},\ldots,\theta_{Mt})$
is the vector of player strengths at time $t$.
Treating the non-strength parameters as fixed and known, or 
by placing priors on them,
models of this form have traditionally been analyzed
via Markov chain Monte Carlo (MCMC) methods.
Examples include the work of
\citet{glickman1993paired}, \citet{glickman1999parameter},
\citet{knorr2000dynamic} and \citet{gorgi2019analysis}.
While MCMC-based approaches are effective in settings with a modest
number  of competitors and time periods, they become computationally infeasible as
the scale of the data increases. 
In environments such as league play or online gaming platforms, 
where thousands of participants compete across many time intervals, 
the dimensionality of the parameter space renders full posterior simulation burdensome. 
Furthermore, in a real-time applications, the primary goal is 
often to estimate current player strength rather than reconstruct historical 
trajectories. 
This practical requirement calls for a more efficient approach that can update ratings 
sequentially as new outcomes are observed, without reprocessing the entire
history of play. 

To address these challenges, we propose in Section~\ref{sec:filter} a 
one-step-ahead filtering algorithm that recursively approximates the posterior 
distributions of player strength using only the most recent game outcomes. 
This approach enables efficient, real-time updates without the need to reprocess 
the full history of results. Similar filtering strategies have been applied in 
simpler paired comparison contexts, 
as in \citet{glickman1999parameter} and \citet{glickman2001dynamic}.

\section{Approximate filtering algorithm for real-time rating updates}\label{sec:filter}

Rather than analyze the full posterior distribution after all game
outcomes in period $t$, we propose a one-step filtering recursion that 
sequentially updates inferences about player strengths, first based on game outcomes, 
and then based on the passage of time.
Let $\bdD_t$ denote the collection of all game outcomes up to and including period $t$.
At the start of period $t$, before game results are observed,
we assume that for each player $i=1,\ldots, M$,
the mean $\mu_{it}$ and variance $\sigma_{it}^2$ of their strength distribution
are known.
We also assume, for now, that the hyperparameters 
$\alpha_0$, $\alpha_1$, $\beta_0$, $\beta_1$ and $\tau^2$ are fixed and known.

Under these assumptions, the prior distribution for $\theta_{it}$ is given by
\begin{equation} \label{eq:prior-t}
    \theta_{it} \mid \bdD_{t-1} \sim \N(\mu_{it}, \sigma_{it}^2).
\end{equation}
To update this prior using the outcomes from period $t$, we apply
a multinomial likelihood based on the probabilities defined
in~(\ref{eq:novel}), yielding an approximate posterior distribution
time period $t$ 
\begin{equation} \label{eq:post-t}
    \theta_{it} \mid \bdD_t \sim \N(\mu_{it}^*, \sigma_{it}^{*2}),
\end{equation}
where $\mu_{it}^*$ and $\sigma_{it}^{*2}$ are the approximate 
posterior mean and variance, respectively.

To obtain the 
prior distribution for period $t+1$, we apply the law of total expectation 
and variance, using the dynamic model in~(\ref{eq:innov}).
For normal densities, this yields
\begin{eqnarray} 
   \label{eq:post-mean}
    \E(\theta_{i,t+1} \mid \bdD_t)
    &=& \E(\E(\theta_{i,t+1} \mid \theta_{it}, \bdD_t) )
    = \E(\theta_{it} \mid \bdD_t) = \mu_{it}^* \; . \\
   \label{eq:post-var}
    \Var(\theta_{i,t+1} \mid \bdD_t)
    &=& \E(\Var(\theta_{i,t+1} \mid \theta_{it}, \bdD_t) ) +
    \Var(\E(\theta_{i,t+1} \mid \theta_{it}, \bdD_t) ) \nonumber \\
    &=& \E(\tau^2  \mid \bdD_t) + \Var(\theta_{it} \mid \bdD_t) = \tau^2 + \sigma_{it}^{*2} .
\end{eqnarray}
Therefore, the prior distribution for period $t+1$ becomes
\begin{equation} \label{eq:prior-t+1}
    \theta_{i,t+1} \mid \bdD_t \sim \N(\mu_{it}^*, \sigma_{it}^{*2} + \tau^2),
\end{equation}
which may then be used to process game results from
period $t+1$, allowing the recursion to proceed forward through time.

It is important to note that, provided the
posterior distribution of $\theta_{it}$ is approximately normal,
the transition from the posterior in~(\ref{eq:post-mean}) 
to the prior in~(\ref{eq:prior-t+1}) represents an exact probabilistic update.
Our overall strategy thus relies on being able to approximate the posterior
density that arises from combining 
the normal prior in~(\ref{eq:prior-t})
with the multinomial likelihood.
This approximation yields the
normal form given in~(\ref{eq:post-t}).
In the sections that follow, we
outline a procedure for computing the approximate 
posterior mean $\mu_{it}^*$ and
variance $\sigma_{it}^{*2}$ for each player $i$.

\subsection{Opponent prior approximation and marginalization}\label{subsec:approx}

To update the prior distribution of $\theta_{it}$ for player $i$
based on game outcomes during period $t$,
we begin by replacing the posterior 
distributions of player $i$'s opponents with their corresponding 
prior distributions.
Letting $-i$ denote the indices of all players other than $i$,
we assume
\begin{equation}\label{eq:prior-approx}
    p(\bdtheta_{-it} \mid \bdD_t) 
    \approx 
    p(\bdtheta_{-it} \mid \bdD_{t-1}). 
\end{equation}
This approximation is justified when player $i$'s opponents have 
well-informed prior estimates of strength and play relatively few games 
during period $t$. 
In such cases, conditioning on the prior rather than the posterior introduces 
minimal information loss and serves as a conservative, and thus robust, simplification. 
A similar approximation was employed in \citet{glickman1999parameter} in the development 
of the Glicko rating system.
With this approximation, the joint posterior distribution
of $\bdtheta_t$ can be approximated as
\begin{eqnarray}
    p(\bdtheta_t \mid \bdD_t) &=&
p(\theta_{it} \mid \bdtheta_{-it},\bdD_t)
p(\bdtheta_{-it} \mid \bdD_t) \nonumber \\
&\approx&
p(\theta_{it} \mid \bdtheta_{-it},\bdD_t)
p(\bdtheta_{-it} \mid \bdD_{t-1}) \nonumber \\
&=&
p(\theta_{it} \mid \bdtheta_{-it},\bdD_t)
\prod_{j\neq i} \N(\theta_{jt} \mid \mu_{jt},\; \sigma_{jt}^2),
\label{eq:approx-joint}
\end{eqnarray}
where $\N(\cdot \mid \mu, \sigma^2)$
denotes a normal density function with mean $\mu$ and 
variance $\sigma^2$.

In~(\ref{eq:approx-joint}), 
we recognize that $\bdD_t = (\bdy_t, \bdD_{t-1})$,
allowing the first factor to be expanded using Bayes' rule,
\begin{eqnarray}
p(\theta_{it} \mid \bdtheta_{-it},\bdD_t)
&\propto&
p(\theta_{it} \mid \bdtheta_{-it},\bdD_{t-1})
p(\bdy_t \mid \bdtheta_t,\bdD_{t-1}) \nonumber \\
&=&
\N(\theta_{it} \mid \mu_{it},\sigma_{it}^2)
p(\bdy_t \mid \bdtheta_t),
\label{eq:conditional-on-opps}
\end{eqnarray}
where the final equality uses the assumption that 
$\theta_{it} \mid \bdtheta_{-it}, \bdD_{t-1} \sim \N(\mu_{it}, \sigma_{it}^2)$,
and that the likelihood depends only on the current period's 
game outcomes and player strengths.

Substituting into~(\ref{eq:approx-joint}) and marginalizing over 
the opponent strengths $\bdtheta_{-it}$, we obtain
\begin{eqnarray}
    p(\theta_{it} \mid \bdD_t) &=&
\int
p(\bdtheta_t \mid \bdD_t)d\bdtheta_{-it}
\approx
\int \left(
p(\theta_{it} \mid \bdtheta_{-it},\bdD_t)
\prod_{j\neq i} \N(\theta_{jt} \mid \mu_{jt},\; \sigma_{jt}^2) \right) d\bdtheta_{-it}
\nonumber \\
&\propto&
\N(\theta_{it} \mid \mu_{it},\sigma_{it}^2)
\int \left(
p(\bdy_t \mid \bdtheta_t)
\prod_{j\neq i} \N(\theta_{jt} \mid \mu_{jt},\; \sigma_{jt}^2) \right) d\bdtheta_{-it}
\nonumber \\
&\propto&
\N(\theta_{it} \mid \mu_{it},\sigma_{it}^2)
\prod_{j \in \Omega_{it}}
\int p(y_{ijt} \mid \theta_{it},\theta_{jt})
\N(\theta_{jt} \mid \mu_{jt},\sigma_{jt}^2) d\theta_{jt},
\label{eq:marginal}
\end{eqnarray}
where $\Omega_{it}$ denotes the index set of opponents of player $i$ during period $t$.
The term
$p(y_{ijt} \mid \theta_{it},\theta_{jt})$
refers
to the game outcome probability model defined in~(\ref{eq:novel}),
expressed as a function of the 
strength parameters, and with dependence on the hyperparameters suppressed for clarity.
The approximation in the first line of~(\ref{eq:marginal}) follows 
from~(\ref{eq:approx-joint}), and the proportionality from the first to the 
second line results from substituting the expression 
in~(\ref{eq:conditional-on-opps}).

Equation~(\ref{eq:marginal})
assumes that each pair of players competes at most once during a given time 
period $t$. 
In practice, however, the same two players may face each other 
multiple times within the same rating period. 
To accommodate this, 
we treat each repeated game as if it were played against 
a distinct opponent having the same prior strength distribution.
This approximation preserves the structural simplicity 
of the computational approach 
while allowing multiple game results against the same opponent to be incorporated. 
It maintains the overall informativeness of the data without introducing statistical 
dependencies that would complicate inference.

\subsection{Gauss-Hermite approximation of multinomial logit integrals}\label{subsec:integral}

The integrals in~(\ref{eq:marginal}), which take the form of normal mixtures of 
multinomial logit probabilities (with non-strength parameters treated 
as known), do not admit straightforward analytic approximations.
In settings where
the likelihood is based on inverse-logit expressions, as 
in~\citet{glickman1999parameter},
a range of approximation techniques have been developed
for evaluating normal mixtures of logistic functions.
These include methods proposed by
\citet{boys1987diagnostic}, 
\citet{crouch1990evaluation},
and \citet{pirjol2013logistic}.
However, when the outcome probabilities follow a multinomial logit model,
as in our case, 
the set of available analytic approximations is considerably more limited.

Several methods are available for approximating normal mixtures of 
multinomial logit probabilities. 
The most direct is to use MCMC simulation; 
however, such iterative techniques are computationally impractical in the 
context of rating large numbers of players, where efficiency is critical. 
As an alternative, closed-form or semi-analytic approximations have been proposed. 
For example, \citet{harding2007using} applied a multivariate Laplace approximation 
to evaluate the required integrals. 
More closely aligned with the approach we adopt below, 
\citet{bhat1995heteroscedastic} and \citet{pryanishnikov2016multinomial} 
approximate the normal mixture integrals using Gaussian quadrature methods, 
which offer a computationally efficient and numerically stable solution for 
problems of this structure.

To approximate the integrals in~(\ref{eq:marginal}), 
we adopt a 2-point Gauss-Hermite quadrature strategy \citep{steen1969gaussian}. 
This method provides a simple yet effective numerical approximation for integrals 
involving normal densities. 
In particular, we approximate the continuous integrals in~(\ref{eq:marginal}) by 
\begin{eqnarray}
\lefteqn{\int p(y_{ijt} \mid \theta_{it},\theta_{jt})
\N(\theta_{jt} \mid \mu_{jt},\sigma_{jt}^2) d\theta_{jt}
\approx} \nonumber \\
&&\frac{1}{2}
p(y_{ijt} \mid \theta_{it},\mu_{jt}-\sigma_{jt})
+
\frac{1}{2}
p(y_{ijt} \mid \theta_{it},\mu_{jt}+\sigma_{jt}),
\label{eq:gh2}
\end{eqnarray}
which corresponds to a symmetric 2-point approximation centered at the mean of 
the normal distribution, with evaluation points located one standard deviation 
above and below the mean. 
% or
% \begin{eqnarray}
% \lefteqn{\int p(y_{ijt} \mid \theta_{it},\theta_{jt})
% \N(\theta_{jt} \mid \mu_{jt},\sigma_{jt}^2) d\theta_{jt}
% \approx}\nonumber \\
% &&\frac{1}{6}
% p(y_{ijt} \mid \theta_{it},\mu_{jt}-\sqrt{3}\sigma_{jt})
% +
% \frac{2}{3}
% p(y_{ijt} \mid \theta_{it},\mu_{jt})
% +
% \frac{1}{6}
% p(y_{ijt} \mid \theta_{it},\mu_{jt}+\sqrt{3}\sigma_{jt}).
% \label{eq:gh3}
% \end{eqnarray}
Using the Gauss-Hermite approximation, the expression in~(\ref{eq:marginal})
becomes
\begin{equation}\label{eq:marginal-gh2-approx}
p(\theta_{it} \mid \bdD_t) 
\propto
\N(\theta_{it} \mid \mu_{it},\sigma_{it}^2)
\prod_{j \in \Omega_{it}}
\left(
\frac{1}{2}
p(y_{ijt} \mid \theta_{it},\mu_{jt}-\sigma_{jt})
+
\frac{1}{2}
p(y_{ijt} \mid \theta_{it},\mu_{jt}+\sigma_{jt})
\right).
\end{equation}
% The expression for a 3-point quadrature is analogous.

Although the approximate posterior for $\theta_{it}$ 
in~(\ref{eq:marginal-gh2-approx}) does not have a closed-form expression, 
it can be evaluated numerically and summarized through standard numerical methods.
However, in the context of a rating system, 
where updates must be performed frequently and efficiently,
it is important to retain both 
computational tractability. 
To this end, we approximate the posterior 
with a normal distribution in closed form by constructing
a Gaussian density that matches the mode and curvature 
of the log-posterior. 
This approach, detailed in the following section, yields a surrogate posterior 
distribution that enables player strengths to be updated in a simple, 
scalable, and analytically convenient manner.

\subsection{Closed-form normal approximation via Newton-Raphson}\label{subsec:normal-density-approx}

The approximating log-posterior density for $\theta_{it}$, based 
on~(\ref{eq:marginal-gh2-approx}), can be written as
\begin{eqnarray}
\lefteqn{
\log p(\theta_{it} \mid \bdD_t) 
=
} \nonumber \\
&&
c + \log \N(\theta_{it} \mid \mu_{it},\sigma_{it}^2)
+
\sum_{j \in \Omega_{it}}
\log \left(
\frac{1}{2}
p(y_{ijt} \mid \theta_{it},\mu_{jt}-\sigma_{jt})
+
\frac{1}{2}
p(y_{ijt} \mid \theta_{it},\mu_{jt}+\sigma_{jt})
\right) \nonumber \\
&=&
c^* - \frac{(\theta_{it}-\mu_{it})^2}{2\sigma_{it}^2}
+
\sum_{j \in \Omega_{it}}
\log \left(
\frac{1}{2}
p(y_{ijt} \mid \theta_{it},\mu_{jt}-\sigma_{jt})
+
\frac{1}{2}
p(y_{ijt} \mid \theta_{it},\mu_{jt}+\sigma_{jt})
\right),
\label{eq:logposterior}
\end{eqnarray}
where $c$ and $c^*$ are constants not involving $\theta_{it}$.
To approximate~(\ref{eq:logposterior}) with the log of a normal density, 
we apply a one-step Newton-Raphson update centered at the prior mean to estimate the 
posterior mean, and approximate the posterior variance using the reciprocal of the 
second derivative of the log-posterior, also evaluated at the prior mean.

% While this approach typically requires iterative optimization, 
% ICCF officials expressed a preference for avoiding such computational overhead. 
% As a result, we adopt a one-step Newton-Raphson update, using the prior mean $\mu_{it}$ 
% as the initial value. 
To avoid the computational cost of full optimization, 
we adopt a one-step Newton-Raphson update centered at the prior mean.
This procedure requires evaluating the first and second derivatives of the log-posterior.
Define
\begin{equation}
u_{ijt}(\theta_{it}) = 
\frac{1}{2}
p(y_{ijt} \mid \theta_{it},\mu_{jt}-\sigma_{jt})
+
\frac{1}{2}
p(y_{ijt} \mid \theta_{it},\mu_{jt}+\sigma_{jt}).
\label{eq:u-def}
\end{equation}
Then the first and second derivatives of the log-posterior with respect to $\theta_{it}$ are
\begin{equation}
\begin{aligned}
\frac{\partial \log p(\theta_{it} \mid \bdD_t)}{\partial \theta_{it}}
&=
\frac{-(\theta_{it}-\mu_{it})}{\sigma_{it}^2}
+
\sum_{j \in \Omega_{it}} 
\frac{
\frac{\partial}{\partial \theta_{it}} u_{ijt}(\theta_{it})
} % numerator
{
u_{ijt}(\theta_{it})
} , % denominator 
\\*[10pt]
\frac{\partial^2 \log p(\theta_{it} \mid \bdD_t)}{\partial \theta_{it}^2}
&=
\frac{-1}{\sigma_{it}^2}
+
\sum_{j \in \Omega_{it}}
\left(
\frac{
u_{ijt}(\theta_{it})
\left( \frac{\partial^2}{\partial \theta_{it}^2}
u_{ijt}(\theta_{it}) \right)
-
\left(\frac{\partial}{\partial \theta_{it}} u_{ijt}(\theta_{it})\right)^2 
} % numerator
{ ( u_{ijt}(\theta_{it}) )^2 
} % denominator
\right) .
\label{eq:derivatives}
\end{aligned}
\end{equation}
The one-step Newton-Raphson update for approximating the posterior mean is then
given by 
\begin{equation}
\mu_{it}^* 
=
\mu_{it} -
\frac{
\frac{\partial \log p(\theta_{it} \mid \bdD_t)}{\partial \theta_{it}}
\Bigr|_{\theta_{it} = \mu_{it}}
}
{
\frac{\partial^2 \log p(\theta_{it} \mid \bdD_t)}{\partial \theta_{it}^2}
\Bigr|_{\theta_{it} = \mu_{it}}
}  .
\label{eq:mean-update}
\end{equation}
The corresponding approximation for the posterior variance is
\begin{equation}
\sigma_{it}^{*2} 
=
\left(
{
\frac{\partial^2 \log p(\theta_{it} \mid \bdD_t)}{\partial \theta_{it}^2}
\Bigr|_{\theta_{it} = \mu_{it}}
}
\right)^{-1} .
\label{eq:var-update}
\end{equation}
This formulation allows posterior updates to be computed independently for each 
player, making the algorithm highly parallelizable and well-suited to 
large-scale implementation.
The computational details for evaluating~(\ref{eq:mean-update}) and~(\ref{eq:var-update})
are provided in Appendix~\ref{app:update},
and the accuracy of the approximation to~(\ref{eq:marginal})
is explored in Appendix~\ref{app:justify}.
While it may be preferable to evaluate~(\ref{eq:var-update}) 
at $\theta_{it}=\mu_{it}^*$ instead of $\theta_{it}=\mu_{it}$, 
the expressions as written allow for computing $\mu_{it}^*$ and 
$\sigma_{it}^*$ in parallel.
This approach, evaluating the posterior variance at
the prior mean, was also adopted in the development of the
Glicko system \citep{glickman1999parameter}.

\subsection{Hyperparameter optimization via predictive likelihood} \label{subsec:hyperparams}

The algorithm described in Sections~\ref{subsec:approx}-\ref{subsec:normal-density-approx}
assumes that the hyperparameters 
$\alpha_0$, $\alpha_1$, $\beta_0$, $\beta_1$ and $\tau^2$ 
are known or have been estimated in advance.
One strategy for estimating these parameters is to 
maximize one-step-ahead predictive accuracy.
This involves fixing the hyperparameters
at candidate values, applying the updating algorithm for all games
through period $t$, and then evaluating predictive performance using 
game outcomes observed in period $t+1$.

Let $L_{kt}$ denote the likelihood contribution of game $k$ during period $t$,
where player $i$ faces player $j$
and the observed outcomes is $y_{ijt}$.
This quantity is defined as
\begin{equation}
\begin{aligned}
L_{kt} &= 
L(\alpha_0,\alpha_1,\beta_0,\beta_1,\tau^2 \mid y_{kt}) \\
&=
\int\int \Pr(Y_{ijt} = y_{kt} \mid
\theta_{it},\theta_{jt},
\alpha_0,\alpha_1,\beta_0,\beta_1)
p(\theta_{it} \mid \bdD_{t-1})p(\theta_{jt} \mid \bdD_{t-1})
d\theta_{it}d\theta_{jt} .
\end{aligned}
\label{eq:integrated-likelihood}
\end{equation}
Because the double integral in~(\ref{eq:integrated-likelihood}) does not
admit a closed-form expression, we approximate it using 
3-point Gauss-Hermite quadrature for each of the marginal distributions
$p(\theta_{it} \mid \bdD_{t-1})$
and
$p(\theta_{jt} \mid \bdD_{t-1})$.
This results in a grid of 9 evaluation points over the joint distribution of 
$(\theta_{it},\theta_{jt})$,
corresponding to all pairwise combinations of three nodes from each marginal. 
We use a 3-point Gauss-Hermite quadrature here to ensure greater numerical precision 
in estimating predictive accuracy. 
This added accuracy is particularly important for model selection, where small 
improvements in fit may influence hyperparameter optimization.
The integral is then approximated as a weighted sum of the likelihood values 
evaluated at these 9 pairs, with weights derived from the product of the 
respective Gauss-Hermite weights. 
This quadrature-based approximation yields an efficient and accurate method for 
evaluating predictive likelihoods, enabling hyperparameter selection via 
out-of-sample validation.

Given a set of candidate hyperparameter values,
the strength updating algorithm is first run through a
designated training period
$t_{\text{train}} < T$.
For each subsequent time period
$t = t_{\text{train}}+1, \ldots, T$, the following 
steps are performed:
\be
\item Compute the predictive log-likelihood for period $t$,
denoted $\ell_t = \sum_k \log L_{kt}$, where the sum is over all games played in
that period.
\item If $t=T$, stop.  Otherwise, proceed to the next step.
\item Apply the strength update algorithm to the game outcomes from period $t$.
\item Increment $t$ by 1 and return to step 1.
\ee
The cumulative predictive log-likelihood, 
\begin{equation}
\sum_{t=t_{\text{train}}+1}^T \ell_t,
\label{eq:cumulative-loglik}
\end{equation}
serves as the objective function for selecting the optimal hyperparameters.

Rather than relying on a fixed grid of candidate hyperparameter values, 
it is common practice to select them adaptively by directly optimizing the 
predictive log-likelihood. 
This can be done using gradient-based optimization methods or, 
alternatively, through derivative-free approaches such as the 
Nelder–Mead simplex algorithm \citep{nelder_simplex_1965}, 
which is available via the {\tt optim} function in R. 
Given the potential multimodality of the objective function, 
it is advisable to run the optimization from multiple initial values to 
improve the chances of identifying a globally optimal solution.

\section{Application to ICCF game outcomes} \label{sec:application}

We demonstrate the proposed updating algorithm
using game outcomes from the ICCF,
based on a dataset of games completed between January 2016 and March 2022. 
Only ``normal'' games were included.
These are defined as games that were completed through play, 
excluding those resolved by default, adjudication, or exceeding time limits. 
The final dataset consisted of a total of 392,658 games involving 8,976 unique players. 
Rating periods were defined in 3-month intervals, 
consistent with the cadence used in the ICCF’s previous Elo-based rating system. 
This partitioning resulted in 25 rating periods over the six-year window.

As part of the evaluation and implementation process, the ICCF was given the option 
to adopt a version of the rating system that incorporated a white-player advantage, 
represented by the parameters $\alpha_0$ and $\alpha_1$. 
In August 2022, the ICCF Congress approved the implementation of the new rating 
algorithm but voted against including a white advantage in the final system. 
This decision reflected concerns about the interpretability and fairness of 
incorporating a white-player advantage, despite its statistical justification.
Accordingly, in the version of the algorithm described below, the parameters $\alpha_0$ 
and $\alpha_1$ are set to zero, reflecting the specification adopted by the ICCF.

\subsection{Hyperparameter tuning and model calibration}\label{subsec:optimal-implementation}

The hyperparameters $\beta_0$, $\beta_1$ and $\tau^2$ were optimized using 
the computational procedure described in Section~\ref{subsec:hyperparams}.
Specifically, the first 20 time periods were treated as training data,
while the remaining 5 periods were used to evaluate the predictive log-likelihood.
The strength parameters were updated sequentially at each 3-month interval
during this evaluation phase.

Prior distributions for player strengths at the start of the time series were 
based on available ICCF ratings.
To facilitate this,
we applied a transformation between ICCF Elo ratings and the model’s 
latent strength parameters that preserved the ratio
$\bdP(y_{ij}=1)/\bdP(y_{ij}=0)$
between players $i$ and $j$. 
Specifically,
for player $i$ at time $t$, the ICCF rating $R_{it}$ was linked to the model-based
mean strength $\mu_{it}$ via
\begin{equation}
R_{it} =
1500 + 
\frac{400}{\log(10)}
\mu_{it}
=
1500+173.72\cdot\mu_{it} \; .
\label{eq:elo-glicko-conversion}
\end{equation}
For players with a rating 
in January 2016, we assumed
a normal prior distribution with a mean $\mu_{it}$ converted from their rating 
based on~(\ref{eq:elo-glicko-conversion})
and a standard deviation of 100 Elo points to reflect moderate uncertainty.
This corresponds to 
$\sigma_{it} = 100/173.72 = 0.576$ on the latent scale.
For unrated players, we specified a default prior mean of 1800 Elo 
and standard deviation of 250, reflecting greater uncertainty in the absence
of historical information. 
These values corresponded to a normal prior for $\theta_{it}$
with $\mu_{it} = 1.727$
and $\sigma_{it} = 1.439$.
We explored alternative specifications for the prior distributions, 
but found that reasonable
variations had minimal impact on the results.

The optimized values of the model hyperparameters, determined
by maximizing the predictive validity criterion described
earlier, are reported in Table~\ref{tbl:optimal-hyperparameters}.
\begin{table}[t]
\centering
\begin{tabular}{r|l}
Parameter & Optimized Value \\ \hline
$\beta_0$ & 0.35338 \\
$\beta_1$ & 0.57041 \\
$\tau$ & 0.46040 \\ \hline   % param_20221018.csv in games-nocolor-optimization-fixed-init-ratings
\end{tabular}
\caption{Optimized values of model hyperparameters.}
\label{tbl:optimal-hyperparameters}
\end{table}
The estimated standard deviation of the innovation process, 
$\tau$, is approximately 0.460 on the logit scale
(i.e., the latent scale on which outcome probabilities are modeled), 
corresponding to a rating volatility of about 80 Elo points over a 3-month period. 
This value reflects the typical degree of
strength fluctuation permitted by the model 
between rating periods. 
The estimated value of $\beta_0$, which controls the baseline 
probability of a draw,
corresponds to a draw rate of 0.416 between two 
evenly matched players rated at 1500 Elo.
In contrast, the estimated $\beta_1$, which captures
the increase in draw likelihood with player strength,
implies a draw probability of 0.950 between two elite players each rated at 2500 Elo,
underscoring the model’s recognition that draws are nearly inevitable at the highest levels. 
Together, these estimates illustrate
the model's capacity to flexibly capture the observed 
variation in draw frequencies across different skill levels.

\begin{figure}[ht]
\centerline{
\includegraphics[width=0.9\textwidth]{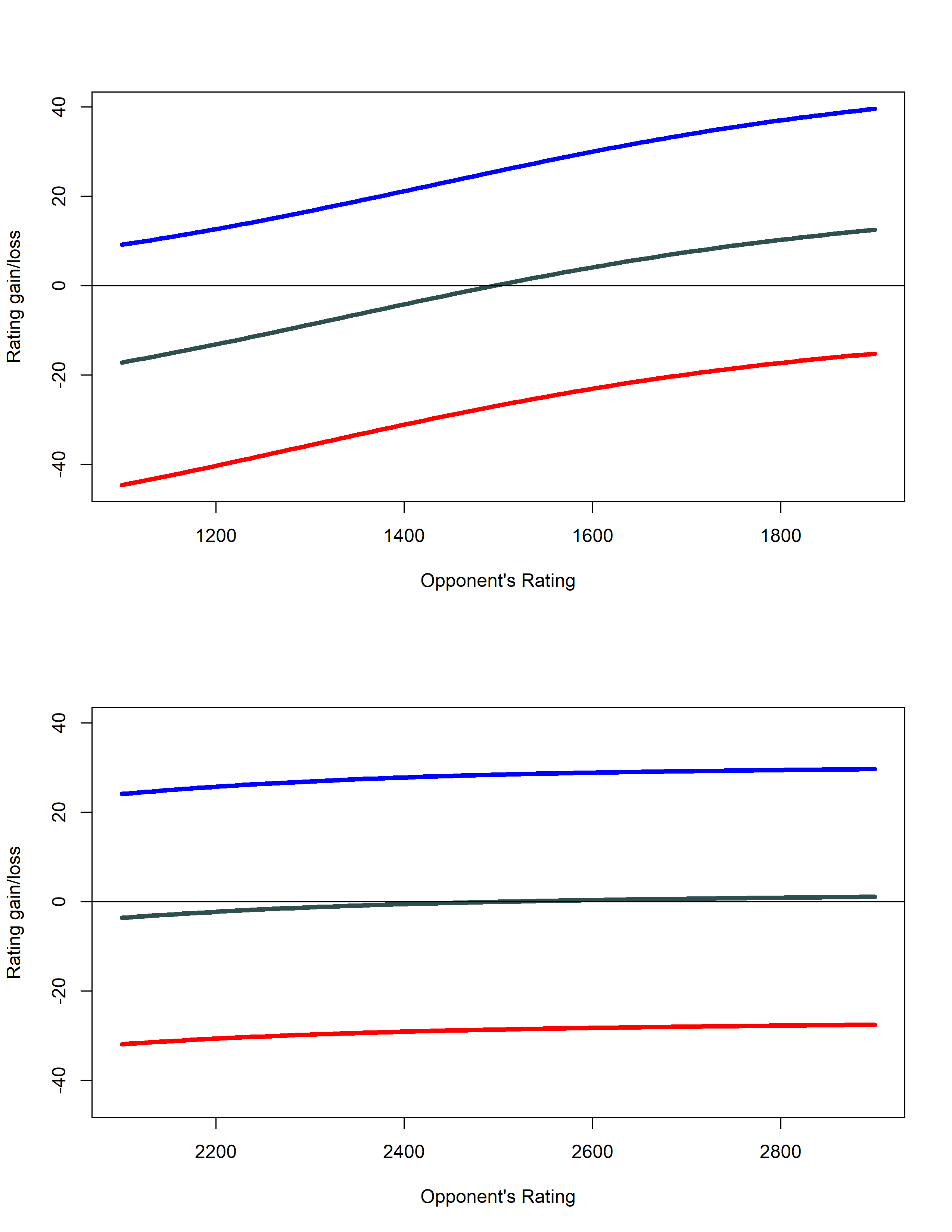}
}
\caption{
Each panel shows the rating change resulting from a win (blue), draw (green), 
or loss (red)
against an opponent with a given Elo rating, 
based on the model parameters in Table~\ref{tbl:optimal-hyperparameters}.
The top panel assumes that the focal player has a rating of 
1500; the bottom panel assumes a rating of 2500.
For all players, the prior standard deviation is assumed to be 100 Elo points.
}
\label{fig:ratchange-optimum}
\end{figure}

The implications of the optimized parameter estimates are
illustrated in Figure~\ref{fig:ratchange-optimum}.
The two panels show how rating changes vary with game outcomes and opponent strength 
under the fitted model. 
In the top panel, where the focal player is rated 1500 Elo, 
rating updates are highly sensitive to opponent rating: 
wins against stronger opponents lead to larger gains, while losses to weaker 
opponents incur greater drops. 
Draws result in modest rating increases when the opponent is stronger and 
modest losses when the opponent is weaker.

In the bottom panel, where the focal player is rated 2500, the magnitude of 
rating changes is much smaller across all outcomes. 
This reflects the model’s tendency to treat results among strong players, 
especially draws, as expected, 
leading to minimal updates. 
In particular, draws yield near-zero rating changes across a wide range 
of opponent ratings, consistent with the model's strength-dependent draw probabilities. 
These patterns demonstrate how the model balances responsiveness to surprising 
results with stability in high-confidence settings.

\subsection{Implementation considerations and ICCF decisions} \label{subsec:practical}

While the hyperparameter values in Table~\ref{tbl:optimal-hyperparameters} 
yielded optimal predictive accuracy,
the resulting Elo-scale ratings were ultimately deemed unsuitable for practical 
use within the ICCF. 
According to feedback from ICCF officials, the ratings produced by the 
optimized system were unlikely to be accepted by the broader ICCF membership.
This led to the adoption of a quasi-optimized version of the model, 
balancing predictive accuracy with the need for interpretability and practical acceptance.

\begin{figure}[ht]
\centerline{
\includegraphics[width=0.9\textwidth]{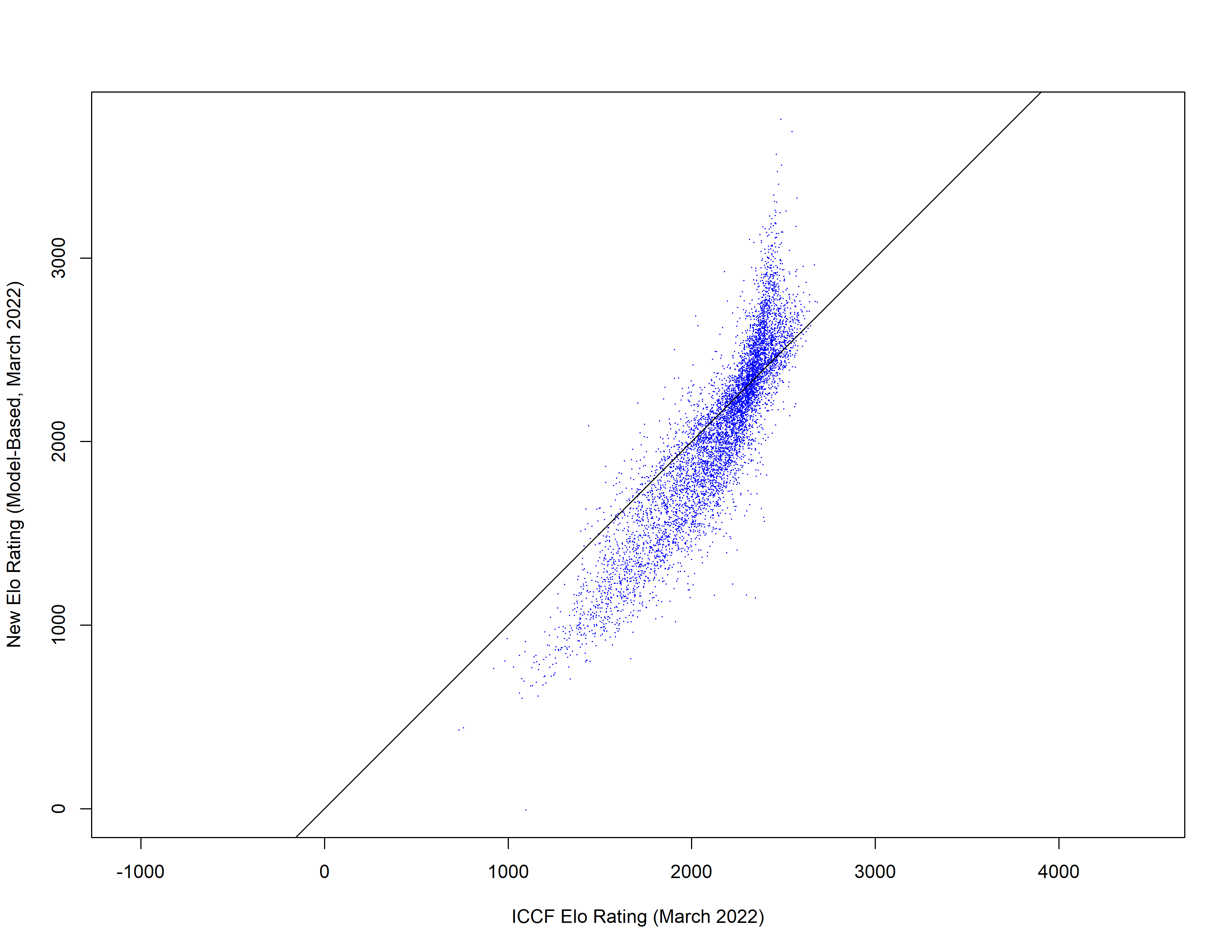}
}
\caption{
Comparison of new ratings in March 2022 based on the optimized hyperparameters in 
Table~\ref{tbl:optimal-hyperparameters} versus official ICCF ratings 
under the Elo-based system. 
The line at $y=x$, indicating perfect agreement, is superimposed for reference.
}
\label{fig:compare-optimum}
\end{figure}

Figure~\ref{fig:compare-optimum} presents a scatter plot 
comparing the 
posterior mean ratings from the optimized model 
(converted to the Elo scale) 
with the official ICCF ratings as of March 2022.
The diagonal 
line $y=x$ is included for reference.
A key issue evident in the figure is that a substantial number of strong players 
with ICCF ratings of 2400 and above would have received ratings exceeding 3000 
under the new system, as seen on the right side of the plot. 
Ratings of this magnitude are extremely rare in other Elo-based systems used for 
slow time-control chess and would likely lack credibility among ICCF members. 
The optimized value of $\tau=0.460$, corresponding to an innovation standard deviation 
of approximately 80 Elo points, resulted in excessive volatility, substantially 
increasing variance from one period’s posterior to the next period’s prior. 
This level of fluctuation was deemed too large for a practical rating system.

Additionally, the optimized values of $\beta_0$ and $\beta_1$ were found to produce 
rating changes that were too small for drawn games between top players, undermining 
a primary motivation for revising the system, and too large for draws between average 
players. 
Together, these issues led to the conclusion that, despite strong predictive 
performance, the optimized system would not be suitable for implementation 
in its current form.

\begin{figure}[ht]
\centerline{
\includegraphics[width=0.9\textwidth]{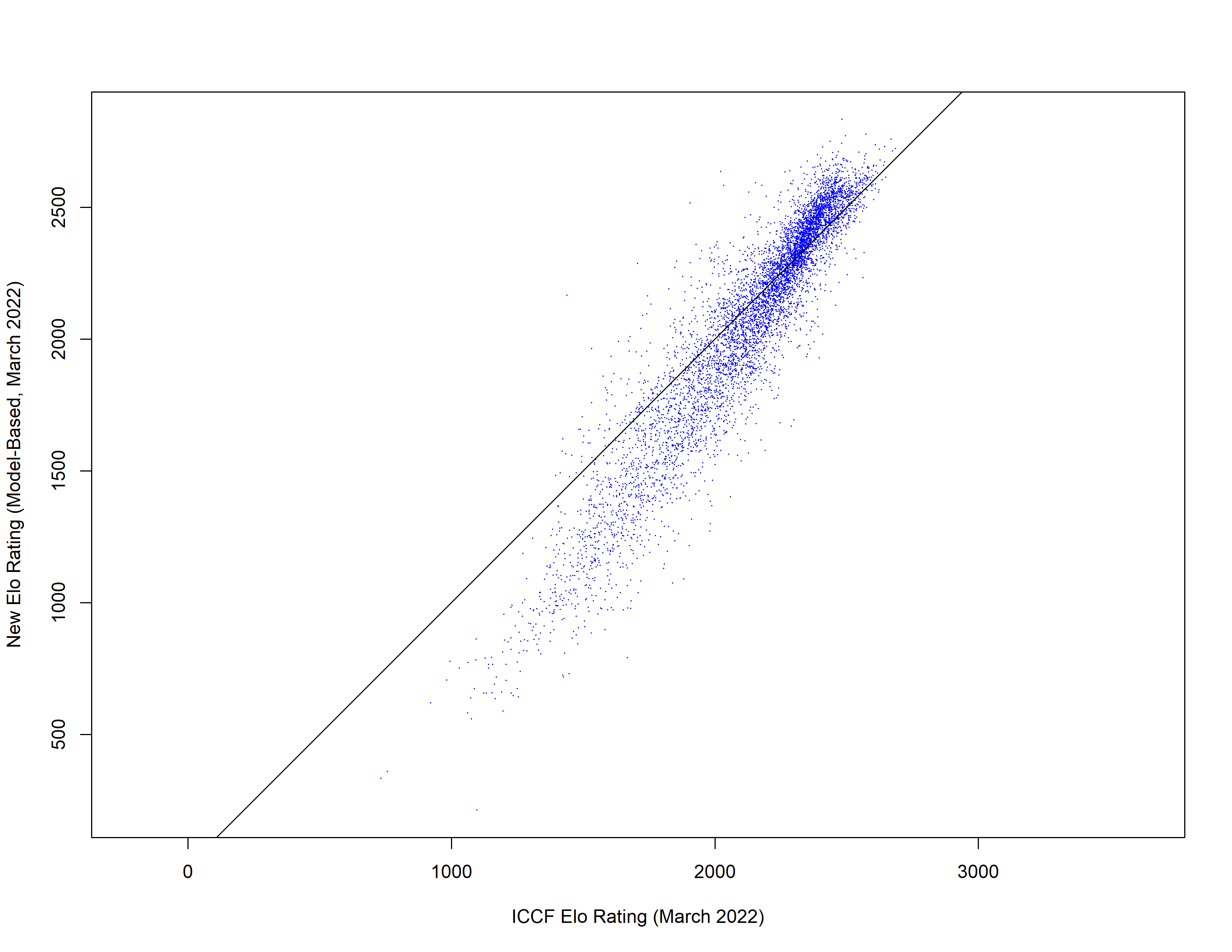}
}
\caption{
Comparison of new ratings in March 2022 using hyperparameters from 
Table~\ref{tbl:quasi-optimal-hyperparameters} versus official ICCF ratings under the 
previous Elo-based system. 
The line at $y=x$, indicating perfect agreement, is included for reference.
}
\label{fig:compare-quasi}
\end{figure}

The rating system ultimately implemented 
by the ICCF used hyperparameter values
shown in Table~\ref{tbl:quasi-optimal-hyperparameters}.
\begin{table}[t]
\centering
\begin{tabular}{r|l}
Parameter & Quasi-optimized value \\ \hline
$\beta_0$ & 1.09861 \\
$\beta_1$ & 0.17037 \\
$\tau$ & 0.14391 \\ \hline   % corresponds to c=25 : 25/173.72
\end{tabular}
\caption{Quasi-optimal values of model hyperparameters used in the 
implemented ICCF rating system.}
\label{tbl:quasi-optimal-hyperparameters}
\end{table}
In contrast to the values in Table~\ref{tbl:optimal-hyperparameters}, 
where $\beta_0$ and $\beta_1$ implied a draw probability of 0.416 between 
two players with strength parameters of 0 (i.e., Elo ratings of 1500), 
and 0.950 between two players with strength parameters of 5.756 
(i.e., Elo ratings of 2500), the quasi-optimized values in 
Table~\ref{tbl:quasi-optimal-hyperparameters} yield more moderate draw probabilities 
of 0.6 and 0.8 for those same player strengths. 
These values produce larger rating changes for top players following drawn games 
and smaller changes for average players—better aligning with ICCF priorities.
Additionally, the lower value of $\tau=0.14391$, corresponding to an innovation 
standard deviation of 25 Elo points, ensures more modest rating fluctuations 
across time periods, leading to a more stable and interpretable rating system. 
The effects of using the quasi-optimized parameters are illustrated in 
Figure~\ref{fig:compare-quasi}, which compares the new ratings (as of March 2022) 
to the ICCF’s legacy Elo-based ratings.
The figure shows that the ratings for top players remain below 3000,
improving face validity relative to the fully optimized system.
While lower-rated players tend to receive lower ratings under the new system 
than under the legacy ICCF system, these outcomes were judged to be consistent 
with expectations and acceptable for implementation.

An additional refinement in the implemented system was a cap on the growth of a 
player’s prior standard deviation from one period to the next. 
Specifically, once a player's uncertainty exceeded a certain threshold, it was no 
longer allowed to increase further. 
The rationale for this rule is that once a player has competed enough to obtain 
a reasonably reliable strength estimate, their uncertainty should not continue 
to grow indefinitely simply due to inactivity.
In practice, the ICCF system was configured to apply the usual innovation variance $\tau^2$
to a player’s posterior variance when computing the prior for the next period only if 
the player’s posterior standard deviation was below 0.691 (equivalent to 120 Elo points). 
If the posterior standard deviation was already at or above this threshold, 
it was carried forward unchanged as the prior standard deviation in the next time period.

\begin{figure}[ht]
\centerline{
\includegraphics[width=0.9\textwidth]{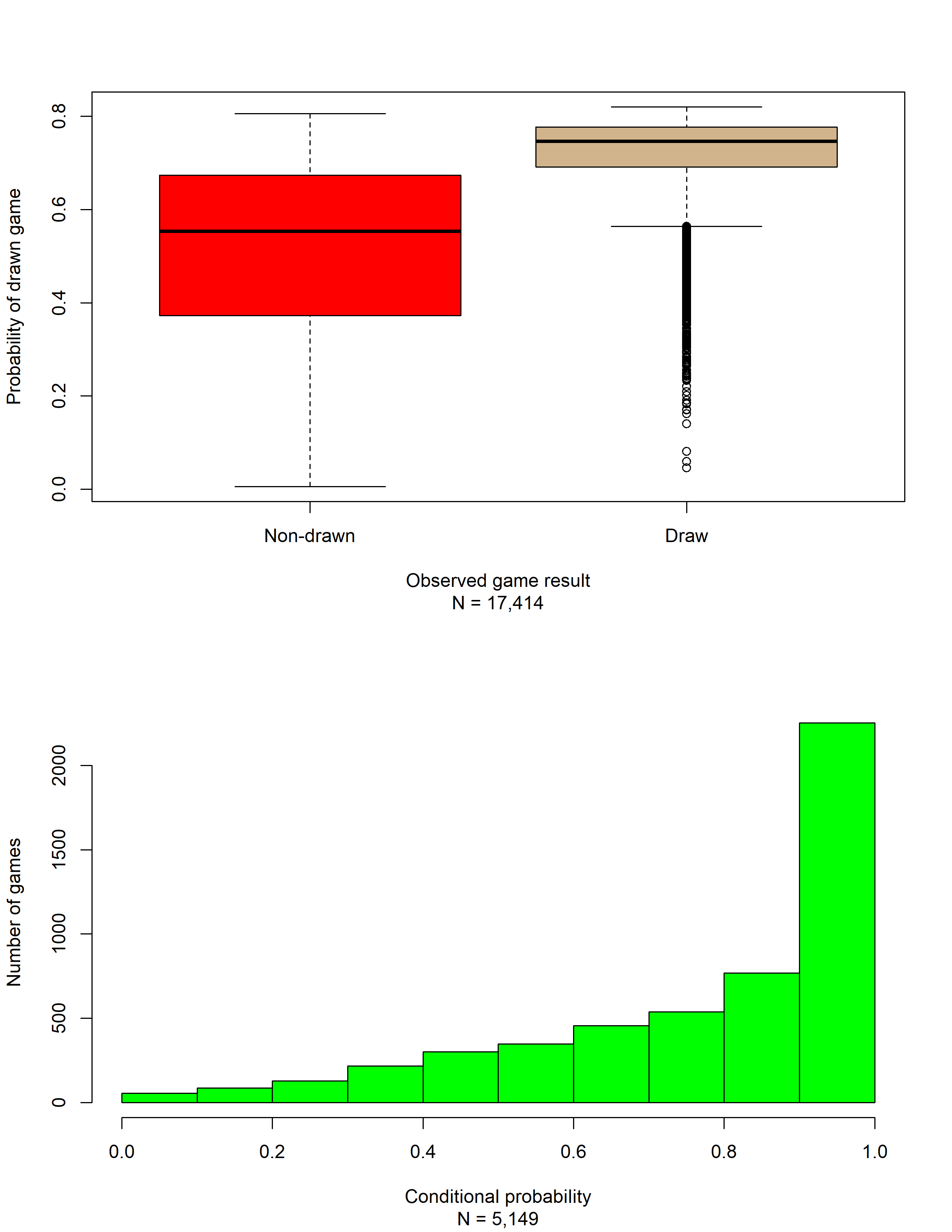}
}
\caption{
Top panel: boxplots of predicted draw probabilities for the 17,414 games in the 
validation sample, separated by whether the game was decisive or drawn. 
Bottom panel: histogram of predicted probabilities of the observed outcome 
(conditional on a decisive result) for the 5,149 decisive games in the validation set. 
All predictions are based on the quasi-optimized hyperparameters in 
Table~\ref{tbl:quasi-optimal-hyperparameters}.
}
\label{fig:quasi-summaries}
\end{figure}

The quasi-optimized hyperparameter values in Table~\ref{tbl:quasi-optimal-hyperparameters}
yield ratings with strong predictive validity, as demonstrated 
in Figure~\ref{fig:quasi-summaries}.
The analysis is based on the five validation periods of ICCF game results. 
The top panel displays pre-game predicted draw probabilities for all 17,414 games 
in the validation sample. 
The left boxplot corresponds to games that ended decisively, 
while the right boxplot corresponds to games that ended in a draw. 
As expected from a well-calibrated model, draw probabilities tend to be higher 
for games that were actually drawn, supporting the validity of the system’s draw predictions.
The bottom panel focuses on the subset of 5,149 games from the validation set that 
resulted in a decisive outcome. 
It shows a histogram of the predicted probabilities assigned to the 
observed (winning) outcome, conditional on the game not being a draw. 
Specifically, these values take the form 
$p = p_{\text win}/(p_{\text win} + p_{\text loss})$,
omitting the draw probability. 
The distribution skews toward higher probabilities, indicating that the model 
generally assigns higher likelihood to outcomes that actually occurred, 
another sign of predictive validity. 
Moreover, only 14.8\% of decisive outcomes were assigned probabilities 
below 0.5, indicating that true upsets were relatively rare.
Together, these analyses provide empirical support for the effectiveness of the 
developed rating system and its ability to generate credible and informative predictions.

\section{Discussion} \label{sec:discuss}

This paper introduced a dynamic rating system for games with 
possible outcomes of a win, loss, or draw, where the probability of a draw is 
modeled as a function of player strength. 
Building on the model developed by \citet{glickman2025paired}, 
the proposed method incorporates a state-space framework in which player abilities 
evolve over time according to a normal random walk. 
The resulting system enables real-time updating of ratings using an approximate 
Bayesian filtering algorithm, ensuring scalability and responsiveness to new results. 
This model was applied to game data from the ICCF,
leading to the development and implementation of a new rating system that 
addresses long-standing issues in handling drawn results, particularly among 
top-level players.

While the implemented system represents a significant step forward, there are 
limitations and opportunities for further refinement. 
Although the ICCF ultimately chose not to implement the system with an 
explicit white advantage, the modeling framework is fully capable of incorporating 
such asymmetries. 
Additionally, the current method assumes a linear relationship between average player 
strength and both the probability of a draw and the advantage of playing white. 
While these forms are convenient and interpretable, 
another possible extension is to introduce nonparametric modeling for draw 
probabilities and white advantage, such as splines or Gaussian processes, 
to better capture complex dependencies observed in real-world data. 
However, such flexibility must be balanced against the need for computational 
efficiency in real-time rating environments.

With its adoption of this system, the ICCF became the first major competitive 
organization to implement a rating algorithm that explicitly models the probability 
of a draw as a distinct, strength-dependent outcome. 
This innovation represents an important evolution in rating methodology, 
particularly in domains where drawn outcomes are prevalent and carry nuanced 
information about player ability. 
The ICCF approved the system in 2022, and it has been in operational use since 2023, 
providing players and officials with a more accurate and context-aware 
evaluation of performance.

\section*{Acknowledgments}
Thanks to Austin Lockwood and the International Correspondence Chess Federation 
for providing the data used in this study, and for their financial support.
Thanks also to Jun Yan for helpful comments on this manuscript.

\addtolength{\baselineskip}{-12pt}
\bibliographystyle{apalike}
\bibliography{refs}

\begin{thebibliography}{}

\bibitem[Bhat, 1995]{bhat1995heteroscedastic}
Bhat, C.~R. (1995).
\newblock A heteroscedastic extreme value model of intercity travel mode choice.
\newblock {\em Transportation Research Part B: Methodological}, 29(6):471--483.

\bibitem[Boys and Dunsmore, 1987]{boys1987diagnostic}
Boys, R. and Dunsmore, I. (1987).
\newblock Diagnostic and sampling models in screening.
\newblock {\em Biometrika}, 74(2):365--374.

\bibitem[Bradley and Terry, 1952]{bradley_rank_1952}
Bradley, R.~A. and Terry, M.~E. (1952).
\newblock Rank analysis of incomplete block designs: {I}. {The} method of paired comparisons.
\newblock {\em Biometrika}, pages 324--345.

\bibitem[Crouch and Spiegelman, 1990]{crouch1990evaluation}
Crouch, E.~A. and Spiegelman, D. (1990).
\newblock The evaluation of integrals of the form $\int_{-\infty}^{\infty} f(t) \exp(-t^2)$: Application to logistic-normal models.
\newblock {\em Journal of the American Statistical Association}, 85(410):464--469.

\bibitem[David, 1988]{david_method_1988}
David, H. (1988).
\newblock {\em The method of paired comparisons}.
\newblock Charles Griffin \& Company, London.

\bibitem[Davidson, 1970]{Davidson1970On}
Davidson, R.~R. (1970).
\newblock On {Extending} the {Bradley}-{Terry} {Model} to {Accommodate} {Ties} in {Paired} {Comparison} {Experiments}.
\newblock {\em Journal of the American Statistical Association}, 65(329):317--328.

\bibitem[Durbin and Koopman, 2012]{durbin2012time}
Durbin, J. and Koopman, S.~J. (2012).
\newblock {\em Time series analysis by state space methods}, volume~38.
\newblock OUP Oxford.

\bibitem[Elo, 1978]{elo1978rating}
Elo, A.~E. (1978).
\newblock {\em The rating of chess players, past and present}.
\newblock Arco Publishing, New York.

\bibitem[Fahrmeir and Tutz, 1994]{fahrmeir1994dynamic}
Fahrmeir, L. and Tutz, G. (1994).
\newblock Dynamic stochastic models for time-dependent ordered paired comparison systems.
\newblock {\em Journal of the American Statistical Association}, 89(428):1438--1449.

\bibitem[Glickman, 1993]{glickman1993paired}
Glickman, M.~E. (1993).
\newblock {\em Paired comparison models with time-varying parameters}.
\newblock Harvard University.

\bibitem[Glickman, 1999]{glickman1999parameter}
Glickman, M.~E. (1999).
\newblock Parameter estimation in large dynamic paired comparison experiments.
\newblock {\em Journal of the Royal Statistical Society Series C: Applied Statistics}, 48(3):377--394.

\bibitem[Glickman, 2001]{glickman2001dynamic}
Glickman, M.~E. (2001).
\newblock Dynamic paired comparison models with stochastic variances.
\newblock {\em Journal of Applied Statistics}, 28(6):673--689.

\bibitem[Glickman, 2025]{glickman2025paired}
Glickman, M.~E. (2025).
\newblock Paired comparison models with strength-dependent ties and order effects.
\newblock {\em arXiv:2505.24783}.

\bibitem[Glickman and Jones, 2024]{glickman2024models}
Glickman, M.~E. and Jones, A.~C. (2024).
\newblock Models and rating systems for head-to-head competition.
\newblock {\em Annual Review of Statistics and Its Application}, 12:259--282.

\bibitem[Gorgi et~al., 2019]{gorgi2019analysis}
Gorgi, P., Koopman, S.~J., and Lit, R. (2019).
\newblock The analysis and forecasting of tennis matches by using a high dimensional dynamic model.
\newblock {\em Journal of the Royal Statistical Society Series A: Statistics in Society}, 182(4):1393--1409.

\bibitem[Harding and Hausman, 2007]{harding2007using}
Harding, M.~C. and Hausman, J. (2007).
\newblock Using a {L}aplace approximation to estimate the random coefficients logit model by nonlinear least squares.
\newblock {\em International Economic Review}, 48(4):1311--1328.

\bibitem[Hastie and Tibshirani, 1986]{hastie1986generalized}
Hastie, T. and Tibshirani, R. (1986).
\newblock Generalized additive models.
\newblock {\em Statistical science}, 1(3):297--310.

\bibitem[Ingram, 2021]{ingram2021extend}
Ingram, M. (2021).
\newblock How to extend {E}lo: a {B}ayesian perspective.
\newblock {\em Journal of Quantitative Analysis in Sports}, 17(3):203--219.

\bibitem[Knorr-Held, 2000]{knorr2000dynamic}
Knorr-Held, L. (2000).
\newblock Dynamic rating of sports teams.
\newblock {\em Journal of the Royal Statistical Society: Series D (The Statistician)}, 49(2):261--276.

\bibitem[Nelder and Mead, 1965]{nelder_simplex_1965}
Nelder, J.~A. and Mead, R. (1965).
\newblock A simplex method for function minimization.
\newblock {\em The Computer Journal}, 7(4):308--313.

\bibitem[Pirjol, 2013]{pirjol2013logistic}
Pirjol, D. (2013).
\newblock The logistic-normal integral and its generalizations.
\newblock {\em Journal of Computational and Applied Mathematics}, 237(1):460--469.

\bibitem[Pryanishnikov and Zigova, 2016]{pryanishnikov2016multinomial}
Pryanishnikov, I. and Zigova, K. (2016).
\newblock Multinomial logit models for the austrian labor market.
\newblock {\em Austrian Journal of Statistics}, 32(4):267--282.

\bibitem[Steen et~al., 1969]{steen1969gaussian}
Steen, N., Byrne, G., and Gelbard, E. (1969).
\newblock Gaussian quadratures for the integrals $\int_0^{\infty} e^{-x^2}f(x) dx$ and $\int_0^b e^{-x^2}f(x) dx$.
\newblock {\em Mathematics of Computation}, 23(107):661--671.

\bibitem[Szczecinski and Djebbi, 2020]{szczecinski2020understanding}
Szczecinski, L. and Djebbi, A. (2020).
\newblock Understanding draws in {E}lo rating algorithm.
\newblock {\em Journal of Quantitative Analysis in Sports}, 16(3):211--220.

\bibitem[West et~al., 1985]{west1985dynamic}
West, M., Harrison, P.~J., and Migon, H.~S. (1985).
\newblock Dynamic generalized linear models and {B}ayesian forecasting.
\newblock {\em Journal of the American Statistical Association}, 80(389):73--83.

\bibitem[Wood, 2011]{wood2011}
Wood, S.~N. (2011).
\newblock Fast stable restricted maximum likelihood and marginal likelihood estimation of semiparametric generalized linear models.
\newblock {\em Journal of the Royal Statistical Society (B)}, 73(1):3--36.

\end{thebibliography}

\addtolength{\baselineskip}{12pt}
\begin{appendices}
\section{Computation of Newton-Raphson updates}\label{app:update}

The one-step Newton-Raphson updates in~(\ref{eq:mean-update})
and~(\ref{eq:var-update})
can be evaluated through a sequence of explicit calculations.
For notational simplicity, we suppress the time index $t$ throughout the
remainder of this section.
Additionally,
conditioning on the hyperparameters is omitted except where their presence is 
relevant to the computations.

Let $p_w$, $p_d$ and $p_\ell$ be the win, draw and loss probabilities
for player $i$, as defined in~(\ref{eq:novel}), 
with the arguments $\theta_i$ and $\theta_j$ suppressed for brevity.
In the expressions for these probabilities, the coefficients multiplying
$\theta_i$ in the exponents can be written as
\begin{equation}
\begin{aligned}
    a_w &= 1 + x_{ij}\alpha_1/8, \\
    a_d &= (1+\beta_1)/2, \\
    a_\ell &= -x_{ij}\alpha_1/8. \\*[6pt]
\end{aligned}
\label{eq:constants-a}
\end{equation}
Define the following expectations over the outcome probabilities:
\begin{equation}
\begin{aligned}
    s_1 &= a_w p_w + a_d p_d + a_\ell p_\ell ,\\
    s_2 &= a_w^2 p_w + a_d^2 p_d + a_\ell^2 p_\ell . \\*[6pt]
\end{aligned}
\label{eq:constants-s}
\end{equation}
The quantity $s_1$ can be interpreted as the expected score
for player $i$, treating $a_w$, $a_d$ and $a_\ell$ as 
outcome-specific scores.
Likewise, $s_2$ represents the expected square score.
For example, when $\alpha_1=0$ % (no strength-dependent advantage to white)
and $\beta_1=0$, % (no strength-dependent draw probability),
the scoring reduces to the conventional values of 1 for a win, 
$\frac{1}{2}$ for a draw, and 0 for a loss.

Under this framework, the first derivatives of the outcome probabilities with respect 
to $\theta_{it}$ are
\begin{equation}
\begin{aligned}
    \frac{\partial p_w}{\partial \theta_i} &=
    p_w(a_w - s_1) ,\\*[10pt]
    \frac{\partial p_d}{\partial \theta_i} &=
    p_d(a_d - s_1) ,\\*[10pt]
    \frac{\partial p_\ell}{\partial \theta_i} &=
    p_\ell(a_\ell - s_1), 
\end{aligned}
\label{eq:first-derivs}
\end{equation}
and the second derivatives are
\begin{equation}
\begin{aligned}
    \frac{\partial^2 p_w}{\partial \theta_i^2} &=
    p_w(a_w^2 - s_2 - 2s_1(a_w - s_1)), \\*[10pt]
    \frac{\partial^2 p_d}{\partial \theta_i^2} &=
    p_d(a_d^2 - s_2 - 2s_1(a_d - s_1)), \\*[10pt]
    \frac{\partial^2 p_\ell}{\partial \theta_i^2} &=
    p_\ell(a_\ell^2 - s_2 - 2s_1(a_\ell - s_1))  .
\end{aligned}
\label{eq:second-derivs}
\end{equation}
These expressions provide the components necessary to compute the first and second 
derivatives of the log-posterior in~(\ref{eq:derivatives}), evaluated at the 
prior mean $\mu_i$, as required by~(\ref{eq:mean-update}) and~(\ref{eq:var-update}).

For each opponent $j$, define the predicted outcome probabilities using the 
Gauss-Hermite quadrature nodes
\begin{equation}
    \begin{aligned}
        P_{wj}^- &= \bdP(Y_{ij}=1 \mid \mu_i, \mu_j-\sigma_j) , \\
        P_{wj}^+ &= \bdP(Y_{ij}=1 \mid \mu_i, \mu_j+\sigma_j) ,\\*[10pt]
        P_{dj}^- &= \bdP(Y_{ij}=\frac{1}{2} \mid \mu_i, \mu_j-\sigma_j) , \\
        P_{dj}^+ &= \bdP(Y_{ij}=\frac{1}{2} \mid \mu_i, \mu_j+\sigma_j) , \\*[10pt]
        P_{\ell j}^- &= \bdP(Y_{ij}=0 \mid \mu_i, \mu_j-\sigma_j) , \\
        P_{\ell j}^+ &= \bdP(Y_{ij}=0 \mid \mu_i, \mu_j+\sigma_j). \\*[10pt]
        \end{aligned}
        \end{equation}
Let $P_j$ be the sum of the predicted probabilities for the observed game outcome:
\begin{equation}
\begin{aligned}
        P_j & = \left\{
        \begin{array}{rl}
            P_{wj}^- + P_{wj}^+ & \mbox{if $y_{ij}=1$,}\\*[6pt]
            P_{dj}^- + P_{dj}^+ & \mbox{if $y_{ij}=\frac{1}{2}$,}\\*[6pt]
            P_{\ell j}^- + P_{\ell j}^+ & \mbox{if $y_{ij}=0$.}
        \end{array} \right.
    \end{aligned}
    \label{eq:pj}
\end{equation}
Also, define the outcome-weighted expectations:
\begin{equation}
    \begin{aligned}
        s_{1j}^- &= a_w P_{wj}^- + a_d P_{dj}^- + a_\ell P_{\ell j}^- ,\\
        s_{1j}^+ &= a_w P_{wj}^+ + a_d P_{dj}^+ + a_\ell P_{\ell j}^+ ,\\
        s_{2j}^- &= a_w^2 P_{wj}^- + a_d^2 P_{dj}^- + a_\ell^2 P_{\ell j}^- ,\\
        s_{2j}^+ &= a_w^2 P_{wj}^+ + a_d^2 P_{dj}^+ + a_\ell^2 P_{\ell j}^+ .\\*[6pt]
    \end{aligned}
\end{equation}
Then the $j$-th term 
in the first derivative sum from~(\ref{eq:derivatives}) is
\begin{equation}
    \delta_{1j} = \left\{
     \begin{array}{rl}
     \left( P_{wj}^-(a_w-s_{1j}^-) + P_{wj}^+(a_w-s_{1j}^+)) \right)/P_j & \mbox{if $y_{ij}=1,$}\\*[6pt] 
     \left( P_{dj}^-(a_d-s_{1j}^-) + P_{dj}^+(a_d-s_{1j}^+)) \right)/P_j & \mbox{if $y_{ij}=\frac{1}{2},$}\\*[6pt]
     \left( P_{\ell j}^-(a_\ell -s_{1j}^-) + P_{\ell j}^+(a_\ell -s_{1j}^+)) \right)/P_j & \mbox{if $y_{ij}=0$.}\\ 
     \end{array}
    \right.
    \label{eq:delta1}
\end{equation}
Note that the factor of $\frac{1}{2}$ in the definition of $u_{ijt}$ 
appears in both the numerator and denominator and cancels out, so it is omitted here.

The corresponding second derivative term is:
\begin{equation}
    \delta_{2j} = \left\{
     \begin{array}{rl}
     \left(
     P_{wj}^-(a_w - s_{2j}^- - 2s_{1j}^-(a_w - s_{1j}^-)) 
     +
     \right.
     & \\*[6pt]
     \left.
     P_{wj}^+(a_w - s_{2j}^+ - 2s_{1j}^+(a_w - s_{1j}^+)) 
     \right)/P_j - \delta_{1j}^2
     & \mbox{if $y_{ij}=1$,} \\*[10pt]
     \left(
     P_{dj}^-(a_d - s_{2j}^- - 2s_{1j}^-(a_d - s_{1j}^-)) 
     +
     \right.
     & \\*[6pt]
     \left.
     P_{dj}^+(a_d - s_{2j}^+ - 2s_{1j}^+(a_d - s_{1j}^+)) 
     \right)/P_j - \delta_{1j}^2
     & \mbox{if $y_{ij}=\frac{1}{2}$,} \\*[10pt]
     \left(
     P_{\ell j}^-(a_\ell - s_{2j}^- - 2s_{1j}^-(a_\ell - s_{1j}^-)) 
     +
     \right.
     & \\*[6pt]
     \left.
     P_{\ell j}^+(a_\ell - s_{2j}^+ - 2s_{1j}^+(a_\ell - s_{1j}^+)) 
     \right)/P_j - \delta_{1j}^2
     & \mbox{if $y_{ij}=0$} .\\*[10pt]
     \end{array}
    \right.
    \label{eq:delta2}
\end{equation}
As before, the factor of $\frac{1}{2}$ 
cancels and is omitted in the expressions above.

With the quantities $\delta_{1j}$ and $\delta_{2j}$ computed as above, 
the posterior mean and variance updates in~(\ref{eq:mean-update}) 
and~(\ref{eq:var-update}) can be rewritten as
\begin{equation}
    \mu_{it}^* = \mu_{it} + 
    \frac{\sum_{j \in \Omega_{it}}  \delta_{1j}}{\sigma_{it}^{-2} - \sum_{j \in \Omega_{it}} \delta_{2j}} ,
    \label{eq:mean-update2}
\end{equation}
\begin{equation}
    \sigma_{it}^{*2} = 
    \frac{1}{\sigma_{it}^{-2} - \sum_{j \in \Omega_{it}} \delta_{2j}} .
    \label{eq:mean-update2}
\end{equation}
These updates are computed independently and in parallel for each player $i$, 
enabling efficient batch processing of all players' posterior distributions.

An artifact of performing the rating updates in parallel is that the draw score value,
$a_d = (1+\beta_1)/2$, 
can introduce  unintended effects when $\beta_1 \neq 0$.
For example, if $\beta_1 > 0$, implying that stronger players are more likely to draw,
then two players with equal prior means who draw a game will both experience an 
increase in their posterior means, according to Equation~(\ref{eq:mean-update2}). 
This behavior is undesirable, as a draw between equally rated players should not 
systematically increase their estimated strengths.
To address this, the implemented system overrides the default value of $a_d$, 
setting it to exactly $\frac{1}{2}$, regardless of the value of $\beta_1$. 
This adjustment ensures that when two players with equal prior means draw a game, 
their posterior means remain unchanged, preserving the symmetry and interpretability 
of the rating update.

\section{Validation of the normal approximation using Gauss-Hermite quadrature}\label{app:justify}

Several numerical approximations are employed to replace the posterior density 
for $\theta_{it}$ in~(\ref{eq:marginal}) with the normal approximation 
defined by the updates in~(\ref{eq:mean-update}) and~(\ref{eq:var-update}).
In this section, we assess the adequacy of this approximation.

Although the distribution of $\theta_{it}$ described
in~(\ref{eq:marginal}) does not admit a closed-form representation,
it can be accurately approximated using Gauss-Hermite quadrature.
Here we focus on the calculation of the posterior mean and variance.
The posterior mean of $\theta_{it}$ is given by the ratio of two nested integrals:
\begin{equation}
\begin{aligned}
\E(\theta_{it} \mid y_{ijt}) 
&=
\frac{
\displaystyle \int_{-\infty}^{\infty} \theta_{it} \cdot \N(\theta_{it} \mid \mu_{it}, \sigma_{it}^2)
\left( \int_{-\infty}^{\infty} p(y_{ijt} \mid \theta_{it}, \theta_{jt}) \cdot \N(\theta_{jt} \mid \mu_{jt}, \sigma_{jt}^2) \, d\theta_{jt} \right)
\, d\theta_{it}
}{
\displaystyle \int_{-\infty}^{\infty} \N(\theta_{it} \mid \mu_{it}, \sigma_{it}^2)
\left( \int_{-\infty}^{\infty} p(y_{ijt} \mid \theta_{it}, \theta_{jt}) \cdot \N(\theta_{jt} \mid \mu_{jt}, \sigma_{jt}^2) \, d\theta_{jt} \right)
\, d\theta_{it}
}  \\*[12pt]
&=
\frac{
\displaystyle \int_{-\infty}^{\infty} 
\int_{-\infty}^{\infty} 
\theta_{it} \cdot 
 p(y_{ijt} \mid \theta_{it}, \theta_{jt}) \cdot 
\N(\theta_{it} \mid \mu_{it}, \sigma_{it}^2)
 \N(\theta_{jt} \mid \mu_{jt}, \sigma_{jt}^2) \, d\theta_{jt} 
\, d\theta_{it}
}{
\displaystyle \int_{-\infty}^{\infty} 
\int_{-\infty}^{\infty} 
p(y_{ijt} \mid \theta_{it}, \theta_{jt}) \cdot 
\N(\theta_{it} \mid \mu_{it}, \sigma_{it}^2)
\N(\theta_{jt} \mid \mu_{jt}, \sigma_{jt}^2) \, d\theta_{jt} 
\, d\theta_{it}
}  .
\end{aligned}
\label{eq:posterior-mean-ratio}
\end{equation}
Similarly, the posterior second moment of $\theta_{it}$ is given by
\begin{equation}
\E(\theta_{it}^2 \mid y_{ijt}) 
=
\frac{
\displaystyle \int_{-\infty}^{\infty} 
\int_{-\infty}^{\infty} 
\theta_{it}^2 \cdot 
 p(y_{ijt} \mid \theta_{it}, \theta_{jt}) \cdot 
\N(\theta_{it} \mid \mu_{it}, \sigma_{it}^2)
 \N(\theta_{jt} \mid \mu_{jt}, \sigma_{jt}^2) \, d\theta_{jt} 
\, d\theta_{it}
}{
\displaystyle \int_{-\infty}^{\infty} 
\int_{-\infty}^{\infty} 
p(y_{ijt} \mid \theta_{it}, \theta_{jt}) \cdot 
\N(\theta_{it} \mid \mu_{it}, \sigma_{it}^2)
\N(\theta_{jt} \mid \mu_{jt}, \sigma_{jt}^2) \, d\theta_{jt} 
\, d\theta_{it}
} ,
\label{eq:posterior-mean-sq-ratio}
\end{equation}
so that
\begin{equation}
    \Var(\theta_{it} \mid y_{ijt}) = \E(\theta_{it}^2) - \left( \E(\theta_{it}) \right)^2.
    \label{eq:posterior-var-ratio}
\end{equation}

To evaluate the expressions in~(\ref{eq:posterior-mean-ratio}) 
and~(\ref{eq:posterior-var-ratio}),
the numerator and denominator can be computed separately through
Gauss-Hermite quadrature.
This method is specifically designed for approximating integrals of the form:
\[
\int_{-\infty}^{\infty} f(z) e^{-z^2} \, dz \approx \sum_{r=1}^{R} w_r f(z_r),
\]
where \( \{z_r\}_{r=1}^R \) are the roots of the degree-\( R \) 
Hermite polynomial \( H_R(z) \), and \( \{w_r\}_{r=1}^R \) are the 
corresponding quadrature weights, given by
\[
w_r = \frac{2^{R-1} R! \sqrt{\pi}}{R^2 (H_{R-1}(z_r))^2}.
\]
To adapt this technique to integrals involving 
normal densities of the form \( \N(\theta \mid \mu, \sigma^2) \), 
we apply the change of variables \( \theta = \mu + \sqrt{2}\sigma z \), 
which yields the quadrature approximation
\[
\int_{-\infty}^{\infty} f(\theta) \cdot \N(\theta \mid \mu, \sigma^2) \, d\theta
\approx \frac{1}{\sqrt{\pi}} \sum_{r=1}^{R} w_r \cdot f(\mu + \sqrt{2} \sigma z_r).
\]
Applying this to both the inner and outer integrals in~(\ref{eq:posterior-mean-ratio}),
we define for each quadrature node 
\( \theta_{it}^{(s)} = \mu_{it} + \sqrt{2} \sigma_{it} z_s \), 
and approximate the inner integrals over $\theta_{jt}$ for opponent \( j \) as
\begin{equation}
I_j(\theta_{it}^{(s)}) \approx \frac{1}{\sqrt{\pi}} \sum_{r=1}^{R} w_r \cdot p(y_{ijt} \mid \theta_{it}^{(s)}, \mu_{jt} + \sqrt{2} \sigma_{jt} z_r).
\label{eq:gh-integral}
\end{equation}
The posterior mean of $\theta_{it}$ can then be approximated as
\begin{equation}
\E[\theta_{it} \mid y_{ijt}] \approx
\frac{
\displaystyle \sum_{s=1}^{R} w_s \cdot \theta_{it}^{(s)} \cdot I_j(\theta_{it}^{(s)})
}{
\displaystyle \sum_{s=1}^{R} w_s \cdot I_j(\theta_{it}^{(s)})
},
\end{equation}
and a similar approximation can be used to compute $\E(\theta_{it}^2 \mid y_{ijt})$,
which in turn yields the posterior variance via~(\ref{eq:posterior-var-ratio}).
These approximations serve as the ground truth for evaluating the accuracy of 
the normal approximation method described in the main text.
In our implementation of the outcome probability model
$p(y_{ijt} \mid \cdot)$, 
we found that using $R=9$ quadrature points provides sufficient accuracy
for evaluating these integrals.

We assessed the accuracy of the normal approximation used in the rating update 
algorithm by comparing it to the more precise Gauss-Hermite quadrature evaluation 
of the posterior density in~(\ref{eq:marginal}). 
To do so, we computed prior means and variances for all players during the final 
period of the ICCF dataset (Q1 2022), using the quasi-optimized hyperparameters 
from Table~\ref{tbl:quasi-optimal-hyperparameters}. 
For each game, we performed single-game posterior updates of the strength parameter 
for the player with white, once using the approximate method described in this paper 
and once using Gauss-Hermite quadrature as a more accurate benchmark.

In both approaches, the same outcome probability model
$p(y_{ijt}\mid \theta_{it}, \theta_{jt})$
and the same  hyperparameters from Table~\ref{tbl:quasi-optimal-hyperparameters}
were used.
These comparisons were conducted separately for all 17,414 games from the validation period.
Table~\ref{tbl:assessment-summaries} summarizes the results of the comparisons.
\begin{table}[ht]
\centering
\begin{tabular}{r|r|rrrr|r|}
 &  & & & & & Log Std Dev \\
 &   &  \multicolumn{4}{c|}{Mean comparisons} & comparions\\
 & $N$ & $\Delta_{\text{approx}}$ & $\Delta_{\text{GH}}$ & $R_{y=x}^2$ & 
 $\Delta_{\text{approx}}$ - $\Delta_{\text{GH}}$
 &  $R_{y=x}^2$\\*[6pt] 
  \hline
All games & 17414 & 0.0402 & 0.0405 & 0.9855 & 0.0076 & 0.9644 \\ 
  Decisive games only & 5149 & 0.1023 & 0.0997 & 0.9912 & 0.0115 & 0.9536 \\ 
  Drawn games only & 12265 & 0.0141 & 0.0157 & 0.9169 & 0.0059 & 0.9765 \\ 
All games, $\mu_1 > 5.06$ & 5842 & 0.0148 & 0.0133 & 0.9623 & 0.0048 & 0.9155 \\ 
  Decisive games, $\mu_1 > 5.06$ & 785 & 0.0654 & 0.0572 & 0.9580 & 0.0094 & 0.8144 \\ 
  Drawn games, $\mu_1 > 5.06$ & 5057 & 0.0070 & 0.0065 & 0.8107 & 0.0041 & 0.9403 \\ 
All games, $3.43 < \mu_1 \leq 5.06$ & 5774 & 0.0261 & 0.0256 & 0.9733 & 0.0063 & 0.9273 \\ 
  Decisive games, $3.43 < \mu_1 \leq 5.06$ & 1278 & 0.0807 & 0.0741 & 0.9797 & 0.0107 & 0.8912 \\ 
  Drawn games, $3.43 < \mu_1 \leq 5.06$ & 4496 & 0.0105 & 0.0118 & 0.8739 & 0.0050 & 0.9445 \\ 
  Games, $\mu_1 \leq 3.43$ & 5798 & 0.0798 & 0.0828 & 0.9874 & 0.0117 & 0.9634 \\ 
  Decisive games, $\mu_1 \leq 3.43$ & 3086 & 0.1205 & 0.1210 & 0.9924 & 0.0124 & 0.9535 \\ 
  Drawn games, $\mu_1 \leq 3.43$ & 2712 & 0.0334 & 0.0393 & 0.9191 & 0.0109 & 0.9784 \\ 
   \hline
\end{tabular}
\caption{
Summaries of single-game updates for 17,414 ICCF games played during 
Q1 2022.
Updates were computed using both the approximate normal method and Gauss-Hermite 
quadrature, under the quasi-optimized hyperparameters from 
Table~\ref{tbl:quasi-optimal-hyperparameters}.
}
\label{tbl:assessment-summaries}
\end{table}
Each row summarizes results for a specific subset of the 17,414 validation games.
Subsets were defined based on two criteria: 
whether the game was decisive or drawn, and the strength of the player with white, 
measured by their prior mean $\mu_1$. 
To capture differences across player strength, 
we divided the games into terciles based on $\mu_1$: 
players with 
$\mu_1 > 5.06$ (strongest), 
$3.43 < \mu_1 \leq 5.06$ (intermediate),
and
$\mu_1 \leq 3.43$ (weakest).
This categorization yielded 12 distinct evaluation subsets.

Each summary in the table reports results for a subset of games, with the following columns:
\begin{itemize}
  \item \( N \): the number of games in the subset.
  \item \( \Delta_{\text{approx}} \): the average absolute change from prior to posterior mean using the approximation method developed in this paper.
  \item \( \Delta_{\text{GH}} \): the average absolute change from prior to posterior mean using Gauss-Hermite quadrature.
  \item \( R^2_{y=x} \): the coefficient of determination from regressing the approximate posterior mean changes against those from Gauss-Hermite quadrature, relative to the identity line \( y = x \).
  \item \( \Delta_{\text{approx}} - \Delta_{\text{GH}} \): the average absolute difference between the posterior mean changes from the approximate method and those from Gauss-Hermite quadrature.
  \item Log Std Dev \( R^2_{y=x} \): the \( R^2 \) statistic comparing the changes in log posterior standard deviations from the approximate method and Gauss-Hermite quadrature, again relative to the line \( y = x \).
\end{itemize}

As shown in Table~\ref{tbl:assessment-summaries}, 
the first two columns,
$\Delta_{\text{approx}}$ and $\Delta_{\text{GH}}$,
are consistently close, indicating strong agreement 
between the approximate method and Gauss-Hermite quadrature in estimating the magnitude of the prior-to-posterior mean updates. 
The accompanying $R^2$ values for the mean changes (column 4) further support this, 
showing that the updates from both methods lie tightly along the identity line $y=x$. 
Notably, the few cases where $R^2<0.95$ correspond to subsets with very small average 
posterior mean changes, as seen in the first two columns, differences that are unlikely 
to be practically significant.

The fifth column, representing the average absolute difference in the prior-to-posterior 
mean changes between the two methods, confirms the numerical closeness of the two 
approaches across all subsets. 
Moreover, the final column demonstrates that the updates to the posterior standard 
deviations (on the log scale) also agree closely between methods.

The only subset where the agreement appears weaker, based on a lower $R^2$ in the 
final column, is for decisive games involving high-strength players. 
This discrepancy likely arises from the relatively small changes in posterior 
standard deviation in such cases, making even minor numerical differences more visible 
in the $R^2$ statistic. 
Nonetheless, the correlation between posterior standard deviations for this subset 
(row 5 of the table) is exceptionally high, at 0.9999913, indicating 
near-perfect agreement between the two methods.

\end{appendices}
\end{document}